\documentclass[submission,copyright,creativecommons]{eptcs}
\providecommand{\event}{DSL 2011}
\usepackage{breakurl}

\makeatletter
\@ifundefined{lhs2tex.lhs2tex.sty.read}%
  {\@namedef{lhs2tex.lhs2tex.sty.read}{}%
   \newcommand\SkipToFmtEnd{}%
   \newcommand\EndFmtInput{}%
   \long\def\SkipToFmtEnd#1\EndFmtInput{}%
  }\SkipToFmtEnd

\newcommand\ReadOnlyOnce[1]{\@ifundefined{#1}{\@namedef{#1}{}}\SkipToFmtEnd}
\usepackage{amstext}
\usepackage{amssymb}
\usepackage{stmaryrd}
\DeclareFontFamily{OT1}{cmtex}{}
\DeclareFontShape{OT1}{cmtex}{m}{n}
  {<5><6><7><8>cmtex8
   <9>cmtex9
   <10><10.95><12><14.4><17.28><20.74><24.88>cmtex10}{}
\DeclareFontShape{OT1}{cmtex}{m}{it}
  {<-> ssub * cmtt/m/it}{}

\DeclareFontShape{OT1}{cmtt}{bx}{n}
  {<5><6><7><8>cmtt8
   <9>cmbtt9
   <10><10.95><12><14.4><17.28><20.74><24.88>cmbtt10}{}
\DeclareFontShape{OT1}{cmtex}{bx}{n}
  {<-> ssub * cmtt/bx/n}{}

\newcommand{\Conid}[1]{\mathit{#1}}
\newcommand{\Varid}[1]{\mathit{#1}}
\newcommand{\anonymous}{\kern0.06em \vbox{\hrule\@width.5em}}

\usepackage{polytable}

\@ifundefined{mathindent}%
  {\newdimen\mathindent\mathindent\leftmargini}%
  {}%

\def\resethooks{%
  \global\let\SaveRestoreHook\empty
  \global\let\ColumnHook\empty}
\newcommand*{\savecolumns}[1][default]%
  {\g@addto@macro\SaveRestoreHook{\savecolumns[#1]}}
\newcommand*{\restorecolumns}[1][default]%
  {\g@addto@macro\SaveRestoreHook{\restorecolumns[#1]}}
\newcommand*{\aligncolumn}[2]%
  {\g@addto@macro\ColumnHook{\column{#1}{#2}}}

\resethooks

\newcommand{\onelinecommentchars}{\quad-{}- }
\newcommand{\commentbeginchars}{\enskip\{-}
\newcommand{\commentendchars}{-\}\enskip}

\newcommand{\visiblecomments}{%
  \let\onelinecomment=\onelinecommentchars
  \let\commentbegin=\commentbeginchars
  \let\commentend=\commentendchars}

\newcommand{\invisiblecomments}{%
  \let\onelinecomment=\empty
  \let\commentbegin=\empty
  \let\commentend=\empty}

\visiblecomments

\newlength{\blanklineskip}
\newcommand{\hsindent}[1]{\quad}
\let\hspre\empty
\let\hspost\empty

\EndFmtInput
\makeatother
\ReadOnlyOnce{polycode.fmt}%
\makeatletter

\newcommand{\hsnewpar}[1]%
  {{\parskip=0pt\parindent=0pt\par\vskip #1\noindent}}

\newcommand{\hscodestyle}{}

\newcommand{\sethscode}[1]%
  {\expandafter\let\expandafter\hscode\csname #1\endcsname
   \expandafter\let\expandafter\endhscode\csname end#1\endcsname}

  {\par\noindent
   \advance\leftskip\mathindent
   \hscodestyle
   \let\\=\@normalcr
   \let\hspre\(\let\hspost\)%
   \pboxed}%
  {\endpboxed\)%
   \par\noindent
   \ignorespacesafterend}

  {\hsnewpar\abovedisplayskip
   \advance\leftskip\mathindent
   \hscodestyle
   \let\hspre\(\let\hspost\)%
   \pboxed}%
  {\endpboxed%
   \hsnewpar\belowdisplayskip
   \ignorespacesafterend}

  {\hsnewpar\abovedisplayskip
   \advance\leftskip\mathindent
   \hscodestyle
   \let\\=\@normalcr
   \(\pboxed}%
  {\endpboxed\)%
   \hsnewpar\belowdisplayskip
   \ignorespacesafterend}

\newcommand{\plainhs}{\sethscode{plainhscode}}

\plainhs

  {\hsnewpar\abovedisplayskip
   \advance\leftskip\mathindent
   \hscodestyle
   \let\\=\@normalcr
   \(\parray}%
  {\endparray\)%
   \hsnewpar\belowdisplayskip
   \ignorespacesafterend}

  {\parray}{\endparray}

  {\(\parray}{\endparray\)}

\def\codeframewidth{\arrayrulewidth}
\RequirePackage{calc}

  {\parskip=\abovedisplayskip\par\noindent
   \hscodestyle
   \arrayrulewidth=\codeframewidth
   \tabular{@{}|p{\linewidth-2\arraycolsep-2\arrayrulewidth-2pt}|@{}}%
   \hline\framedhslinecorrect\\{-1.5ex}%
   \let\endoflinesave=\\
   \let\\=\@normalcr
   \(\pboxed}%
  {\endpboxed\)%
   \framedhslinecorrect\endoflinesave{.5ex}\hline
   \endtabular
   \parskip=\belowdisplayskip\par\noindent
   \ignorespacesafterend}

\newcommand{\framedhslinecorrect}[2]%
  {#1[#2]}

  {\(\def\column##1##2{}%
   \let\>\undefined\let\<\undefined\let\\\undefined
   \newcommand\>[1][]{}\newcommand\<[1][]{}\newcommand\\[1][]{}%
   \def\fromto##1##2##3{##3}%
   }{\) }%

  {\let\orighscode=\hscode
   \let\origendhscode=\endhscode
   \def\endhscode{\def\hscode{\endgroup\def\@currenvir{hscode}\\}\begingroup}
   \orighscode\def\hscode{\endgroup\def\@currenvir{hscode}}}%
  {\origendhscode
   \global\let\hscode=\orighscode
   \global\let\endhscode=\origendhscode}%

\makeatother
\EndFmtInput

\title{Efficient and Correct Stencil Computation via Pattern Matching and
  Static Typing}

\author{Dominic Orchard 
\institute{Computer Laboratory, University of Cambridge, UK}
\email{dominic.orchard\ensuremath{\mathord{@}}cl.cam.ac.uk}
\and
Alan Mycroft
\institute{Computer Laboratory, University of Cambridge, UK}
\email{am\ensuremath{\mathord{@}}cl.cam.ac.uk}
}

\def\titlerunning{Efficient and Correct Stencil Computations
  via Pattern Matching and Static Typing}
\def\authorrunning{D. A. Orchard, A. Mycroft}

\usepackage{fancyvrb}
\usepackage{amsmath}
\usepackage{amssymb}
\usepackage{enumitem}
\usepackage{subfigure}
\usepackage{stmaryrd}

\usepackage{graphics}
\usepackage{color}

\newcommand{\myfigref}[1]{Figure \ref{#1}}

\newcommand{\ttt}[1]{\texttt{#1}}
\newcommand{\interp}[1]{\llbracket{#1}\rrbracket}

\RecustomVerbatimEnvironment{Verbatim}{Verbatim}{fontsize=\footnotesize,xleftmargin=0.5cm,xrightmargin=0.5cm,frame=lines,framesep=3mm,commandchars=\\\{\}}
\CustomVerbatimEnvironment{LLVerbatim}{Verbatim}{fontsize=\small,xleftmargin=0.5cm,xrightmargin=0.5cm,frame=lines,framesep=3mm,commandchars=\\\{\}}
\newcommand{\at}{\texttt{\ensuremath{\mathord{@}}}$\,$}

\begin{document}
\maketitle

\begin{abstract}
Stencil computations, involving operations over the elements of an array,
are a common programming pattern in scientific computing, games, and
image processing.
As a programming pattern, stencil computations are highly
regular and amenable to optimisation and parallelisation.
However, general-purpose languages obscure this regular pattern from the
compiler, and even the programmer, preventing optimisation and
obfuscating (in)correctness.
This paper furthers our work on the \emph{Ypnos} domain-specific
language for stencil computations embedded in Haskell.
Ypnos allows declarative, abstract specification of 
stencil computations, exposing the structure of a problem to the
compiler and to the programmer via specialised syntax.
In this paper we show the decidable safety guarantee that
well-formed, well-typed Ypnos programs cannot index outside of array
boundaries. Thus indexing in Ypnos is safe and run-time
bounds checking can be eliminated.
Program information is encoded as types,
using the advanced type-system features of the Glasgow
Haskell Compiler, with the safe-indexing invariant enforced at compile time via
type checking.
\end{abstract}

\section{Introduction}

\emph{Stencil computations}, otherwise known as \emph{stencil codes}
\cite[p221]{yang2006high} or \emph{structured grid computations}
\cite{berkeley06}, are a ubiquitous pattern in programming, particularly
in scientific computing, games, image processing, and similar
applications. Stencil computations are characterised by 
array operations where the elements of an output array are computed from 
corresponding elements, and their surrounding neighbourhood of elements, in
an input array, or arrays. An example is the \emph{discrete
  Laplace} operator which, for two-dimensional problems, can be
written in C as the following, for input array \texttt{A} and output
array \texttt{B}:
\begin{Verbatim}
  for (int i=1; i<(N+1); i++) \{
    for (int j=1; j<(M+1); j++) \{
      B[i][j] = A[i+1][j] + A[i-1][j] + A[i][j+1] + A[i][j-1] - 4*A[i][j]
    \}
  \}
\end{Verbatim}
where \texttt{A} and \texttt{B} are arrays of size $(N+2) \times (M+2)$ with
indices ranging from $(0, 0)$ to $(N+1, M+1)$. The \emph{iteration space}
of the \ttt{for}-loops ranges from $(1, 1)$ to $(N, M)$ where the
elements outside of the iteration space provide a \emph{halo} of the
\emph{boundary conditions} for the operation. The discrete Laplace operator is
an example of a \emph{convolution} operation in image processing
\cite{machine-vision}.

Indexing in C is unsafe: any indexing operation may,
without any static warnings, address memory ``outside'' the
allocated memory region for \texttt{A} or \texttt{B}, causing a
program crash or, even worse: affecting the numerical
accuracy of the algorithm without crashing the program. Bounds-checked 
array access is provided by many languages and libraries
to prevent unsafe program behaviour due to out-of-bounds access.
However, bounds checking every array access has considerable
performance overhead \cite{markstein1982optimization}; we measured an overhead of
roughly $20\%$ per iteration for safe-indexing over unsafe-indexing in
Haskell for a two-dimensional discrete Laplace operator on a $512\times512$
image (see Appendix \ref{experiments} for details).

If it is known, or can be proved, that memory accesses are within
the bounds of allocated regions, then costly bounds checking can be safely 
eliminated. In the presence of \emph{general}, or \emph{random},
indexing, such as in most language and libraries, automatic proof of
the absence of out-of-bounds access is in general undecidable as
indices may be arbitrary expressions.



This paper presents our latest work on \emph{Ypnos}, an \emph{embedded
domain-specific language} in Haskell. Ypnos programs express abstract,
declarative specifications of stencil computations which are
guaranteed free from out-of-bounds array access, and thus 
array indexing operations without bounds checking can be used by the
implementation. Ypnos lacks general array indexing but instead
provides a form of relative indexing via a novel syntactic
construction called a \emph{grid pattern}, as introduced in
\cite{ref-ypnos}. Grid patterns provide decidable compile-time
information about array access.

Our previous paper said little about boundary conditions and safety,
instead focussing on expressivity and parallelisation.
In this paper, we introduce language constructs in Ypnos for specifying
the boundary conditions of a problem and the mechanism by which
array indexing is guaranteed safe. 
Grid patterns are a typed construction which
encode array access information in the types. Safe indexing
is enforced via type checking at compile time, facilitated
by Haskell type system features such as \emph{type classes}
\cite{hall1996type, haskell98} and more advanced type system features
found in the Glasgow Haskell Compiler (GHC) including:
\emph{generalised algebraic data types} (GADTs)
\cite{jones2004wobbly, peyton2006simple} and \emph{type families}
\cite{associated-type-synonyms}. Dependent types have been used
to enforce safe indexing in more general settings
\cite{swierstra2008dependent, xi1999dependent}. However, a
well-developed dependent type system is not required to
enforce safe indexing in Ypnos due to the relative indexing scheme provided
by the grid pattern syntax; instead GADTs suffice.

This paper was inspired by related work on the Repa array
library for Haskell \cite{repa1}. The most recent paper
\cite{repa2} introduces a mechanism for the abstract specification of
convolution masks and a data structure for describing stencils.
Given an array with adequate boundary values a stencil application
function can safely elide bounds checking. We believe that the Ypnos
offers greater flexibility and expressivity, particularly for
higher-dimensionality problems and complex boundary
conditions, facilitated by \emph{grid patterns} and Ypnos's approach
to boundaries and safety (see Section
\ref{related-work} for more on Repa).





Section \ref{ypnos} introduces the core aspects of the Ypnos
language relevant to an end-user. We do not discuss all
Ypnos features here. In particular we elide reductions, iterated
stencil computations, and automatic parallelisation, details of
which can be found in \cite{ref-ypnos}.
Section \ref{types} discusses details of the Ypnos
implementation, in particular the type-level techniques
used for encoding and enforcing safety. The desugaring of Ypnos's
specialised syntax is also informally described in this
section. Section \ref{results}
provides a quantitative analysis of the effectiveness of Ypnos for
producing correct, efficient stencil programs, along with performance
numbers. Section \ref{related-work} considers related work of other DSLs and
languages for array programming. Section \ref{further-work} discusses
further work and Section \ref{conclusion} concludes with a discussion
on the relevance of the techniques presented in this paper to the
wider DSL audience. Appendix \ref{proof} provides a proof sketch of
the soundness of the approach to safety presented in this paper.

The Ypnos implementation, and the source code for the programs used in
this paper, can be downloaded from \url{http://github.com/dorchard/ypnos}.

A deep knowledge of Haskell is not required to understand the
paper, although knowledge of ML or other functional
language is helpful. A brief introduction to the more advanced 
GHC/Haskell type system features employed in the implementation is
provided in Appendix \ref{haskell-types}. Throughout, types starting with a
lower-case letter are \emph{type variables} and are implicitly
universally quantified; types starting with an upper-case letter are
\emph{type constructors}, which are (possibly nullary) constructors
e.g. \ensuremath{\Conid{Int}}.

\section{Introduction to Ypnos}
\label{ypnos}

We introduce the main aspects of Ypnos that would be
relevant to an end-user, using the example of the two-dimensional
discrete Laplace operator:
\begin{Verbatim}
[dimensions| X, Y |]

laplace2D = [fun|  X*Y:| _   t  _ |
                       | l {\at}c  r |
                       | _   b  _ | -> t + l + r + b - 4.0*c  |]

laplaceBoundary = [boundary| Double from (-1, -1) to (+1, +1) -> 0.0 |]

grid = listGrid (Dim X :* Dim Y) (0, 0) (w, h) img_data laplaceBoundary
grid' = runA grid laplace2D
\end{Verbatim}
Here \ttt{img\_data}, \ttt{w}, and \ttt{h} are free variables defined
elsewhere, providing the data of a grid as a list of double-precision floating point
values, and the horizontal and vertical size of the grid respectively.

Ypnos mostly comprises library functions thus the syntax of Ypnos
programs is largely that of Haskell. However there is some
important Ypnos-specific syntax written inside of \emph{quasiquoting}
brackets \cite{quasiquotes} and expanded by macro functions, with the forms:
\begin{Verbatim}
[dimension| {\emph{...ypnos code...}} |]
[fun| {\emph{...ypnos code...}} |]
[boundary| {\emph{...ypnos code...}} |]
\end{Verbatim}
The macros essentially give an executable semantics to the specialised
syntax of Ypnos (discussed further in Section \ref{types}).
%
Within the quasiquoted brackets any expression following an arrow symbol
\ttt{->} is a Haskell expression i.e. in Lisp terminology, expressions
following \ttt{->} are \emph{backquoted}. 

\subsection{Grids and Dimensions}

Ypnos is built around a central data type of $n$-dimensional
immutable arrays. The term \emph{grid} is used instead of \emph{array}
to escape implementational connotations. There are a number of grid
constructors, one of which is used in the above example:
\ttt{listGrid}, which takes five arguments: a
\emph{dimension term}, the lower extent of the grid, the upper extent
of the grid, a list of element values, and a structure describing the
grid's \emph{boundary behaviour}.

Dimension terms define the \emph{dimensionality} of a grid,
naming the dimensions. Any dimension names used in a program must first be
declared in a single declaration via the \ttt{[dimension| ... |]}
macro, e.g. in the example the \ttt{X} and \ttt{Y} dimensions are
declared. A dimension term is constructed from one or more dimension
identifiers, prefixed by \ttt{Dim} constructors, where many dimension
can be composed by the binary \ttt{:*} dimension-tensor operation.

\subsection{Indexing: Grid Patterns}

Ypnos has no general indexing operations on grids. Indexing is instead
provided by a special pattern matching syntax called a \emph{grid pattern}, first
introduced in the earlier Ypnos paper \cite{ref-ypnos}.
A grid pattern is a group of variable or wildcard patterns 
which are matched onto a subset of elements in a grid
relative to a particular element. For example, the following is a
one-dimensional grid pattern on the $X$ dimension:
\begin{LLVerbatim}
X:| l {\at}c r |
\end{LLVerbatim}
This grid pattern, which matches on a one-dimensional grid of
dimension \ttt{X}, binds the variable pattern \ttt{c} to the element of the grid
which is indexed by a \emph{cursor} index internal to the grid data
structure. The cursor is the index of the current iteration of an
array operation. The variable patterns \ttt{l} and \ttt{r} are bound
to elements relative to the cursor element i.e. \ttt{l} to the
preceding element and \ttt{r} to the succeeding element. The above
grid pattern is analogous to the following bindings in C: \ttt{l = A[i-1]; c =
  A[i]; r = A[i+1];} for an array \ttt{A} and index \ttt{i}, where
\ttt{i} is the \emph{cursor} index.

Every grid pattern must contain exactly one sub-pattern prefixed by
\ensuremath{\mathord{@}} which is matched to the grid element indexed by the grid's cursor.
The relative lexical position of the remaining patterns to the
cursor pattern determines to which element in the grid each pattern
should be matched.  

One-dimensional grid patterns can be nested to provide $n$-dimensional
patterns. For example, a pattern match on a two-dimensional grid, of dimensionality
\ttt{Dim X :* Dim Y}, can be defined:
\begin{LLVerbatim}
 Y:|   X:| lt {\at}lc lb |
    {\at} X:| ct {\at}cc cb |
       X:| rt {\at}rc rb | |
\end{LLVerbatim}
Here the outer grid pattern has exactly one pattern prefixed by \ensuremath{\mathord{@}}
and the inner grid patterns also have exactly one pattern marked by
\ensuremath{\mathord{@}}. 
Grid patterns are prefixed by a dimension term to disambiguate the
dimension being matched upon by a stencil.
For syntactic convenience, Ypnos provides a two-dimensional
grid pattern syntax which is considerably easier to read and write than
nested patterns. In two-dimensional grid pattern
syntax, the above pattern can be written:
\begin{LLVerbatim}
 X*Y:|  lt   lc  lb |
     |  ct {\at}cc  cb |
     |  rt   rc  rb |
\end{LLVerbatim}
The layout of the grid pattern syntax is essentially pictorial in its
representation of a stencil's access pattern, where the spatial
relationships of patterns match the spatial relationships of data.

The above access pattern is known as a \emph{nine-point}
two-dimensional stencil. The Laplace example uses a \emph{five-point}
stencil where a wildcard pattern is used to ignore the corner
elements. 

Grid patterns restrict the programmer to relative array-indexing,
expressed as a static construction that requires no
evaluation or reduction. Grid patterns thus provide decidable
compile-time information about the access pattern of a program without
the need for analysis further than parsing. Because indexing is
static, the access pattern of a stencil function can be encoded as a
type (discussion further in Section \ref{types}).
The \ttt{fun} macro allows a \emph{stencil function} to be
defined by a grid pattern and, following the arrow symbol \ttt{->}, a Haskell
expression which defines the body of the stencil function.


\subsection{Applying Stencil Functions and Boundaries}
\label{boundaries-intro}

There are several functions for applying stencil functions to
grids. The example uses \ttt{runA} which applies a stencil
function at each position in the parameter grid by instantiating the internal
cursor index of the grid to each index in the range $(0, 0)$
to $(\texttt{w}-1, \texttt{h}-1)$ (inclusive), writing the results
into a new grid. This range of indices is called the \emph{extent} of the grid.

The grid pattern of \ttt{laplace2D} accesses 
indices $(i-1, j)$, $(i+1, j)$, $(i, j-1)$, and $(i, j+1)$ relative to
the cursor index $(i, j)$ thus application of the stencil function
at indices around the \emph{edge} of the grid, e.g. $(0, 0)$, results in
out-of-bounds access. Errors or undefined behaviour are
avoided by specification of boundary behaviour by 
\ttt{laplaceBound}, which is passed to the \ttt{listGrid}
constructor.

\begin{figure}[h]
\centering
\begin{minipage}{0.3\linewidth}
\scalebox{0.33}{\includegraphics{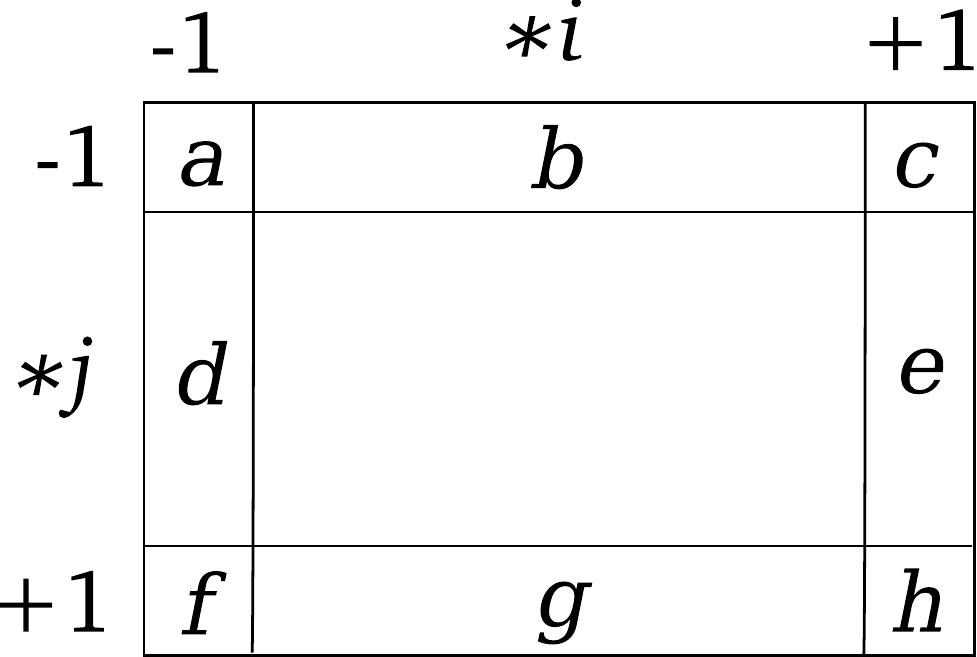}}
\end{minipage}
\begin{minipage}{0.3\linewidth}
\vspace{-1em}
\begin{align*}
\begin{array}{rl}
a = \ttt{(-1, -1)} & e = \ttt{(+1, *j)} \\
b = \ttt{(*i, -1)} & f = \ttt{(-1, +1)} \\
c = \ttt{(+1, -1)} & g = \ttt{(*i, +1)} \\
d = \ttt{(-1, *j)} & h = \ttt{(+1, +1)}
\end{array}
\end{align*}
\end{minipage}
\vspace{1em}
\hrule
\caption{Boundary region descriptors for a two-dimensional grid with a one element-deep boundary}
\label{boundary-regions}
\end{figure}

Boundary behaviour can be specified in a number of ways inside the
\texttt{boundary} quasiquoted macro. In the example we defined the
boundary behaviour by:
\begin{Verbatim}
[boundary| Double from (-1, -1) to (+1, +1) -> 0.0 |]
\end{Verbatim}
which specifies a one-element deep \emph{boundary
  region}, or \emph{halo}, around a grid of element type \ttt{Double}
with a default value of \ttt{0.0}. The
\ttt{from ... to ...} syntax is itself
short-hand for a more verbose description specifying smaller 
sub-regions of the boundary. The above is short-hand for the following:
\begin{Verbatim}
[boundary| Double  (-1, -1) -> 0.0
                   (*i, -1) -> 0.0
                   (+1, -1) -> 0.0
                   (-1, *j) -> 0.0
                   (+1, *j) -> 0.0
                   (-1, +1) -> 0.0
                   (*i, +1) -> 0.0
                   (+1, +1) -> 0.0 |]
\end{Verbatim}
Each case defines the value of a boundary region where the left-hand
side of \ttt{->} is a \emph{region descriptor}. \myfigref{boundary-regions}
gives a pictorial representation of boundary regions, and their
descriptors, for a two-dimensional grid with a one-element boundary
region. Note, \texttt{*v} denotes a variable binding for a variable \texttt{v}.

\noindent
The grammar of boundary terms is the following, where $n$ are
natural numbers greater than one, $e$ are Haskell expressions, and $v$ are variables:
\begin{align*}
B \; ::= & \;\; \texttt{from} \;\; I_1 \;\; \texttt{to} \;\; I_2 \;\;
 \texttt{->} \;\; e \;\; | \;\; I \;\; \texttt{->} \;\; e
 \;\; | \;\; I \;\; g \;\; \texttt{->} \;\; e \;\; \\
I \; ::= & \;\; P \;\; | \;\; \texttt{($P$, $P$)} \;\; | \;\; \texttt{($P$,
  $P$, $P$)} \;\; | \;\; \texttt{($P$, $P$, $P$, $P$)} \;\; | \;\; \ldots
\\
P \; ::= & \;\; \texttt{-}n \;\; | \;\; \texttt{+}n \;\; | \;\; \texttt{*}v 
\end{align*}
A boundary definition consists of an explicit element type and a
number of $B$ terms which specify
either a range of boundary regions with the same value for the elements
of each region or a specific boundary region. The three forms are called the \emph{range form}, the \emph{specific
  form}, and the \emph{specific parameterised form}.
The terminal $\ttt{+}n$ denotes a boundary region $n$
elements after the upper extent of a grid dimension; $\ttt{-}n$
denotes a boundary region $n$ elements before the lower extent of a
grid dimension; $\ttt{*}v$ denotes any index inside the extent of a
grid dimension where the value of the index is bound to $v$ (see
\myfigref{boundary-regions}).

As an example of a more complicated boundary definition, the following
specifies the boundary of a two-dimensional grid where the ``left''
side of the grid (i.e. regions $a$, $d$, and $f$ in
\myfigref{boundary-regions}) has default value $1.0$ to a depth of
one element, the ``right'' side (i.e. regions $c$, $e$, and $h$ in
\myfigref{boundary-regions}) has default value $2.0$, one-element deep, and the
remaining edges have default value $0.0$, one-element deep:
\begin{Verbatim}
[boundary| Double  from (-1, -1) to (-1, +1) -> 1.0
                   from (+1, -1) to (+1, +1) -> 2.0
                   (*i, -1) -> 0.0
                   (*i, +1) -> 0.0 |]
\end{Verbatim}
More complex boundary behaviour such as \emph{reflecting} or
\emph{wrapping} boundaries can be described by the \emph{specific
  parameterised form} which is the same as the \emph{specific form}
with an additional parameter for the grid for which the boundary is
being defined. As an example of the specific parameterised form, the
following defines the boundary of a two-dimensional grid where the
``top'' edge reflects the values inside the grid to a depth of one
element, the ``left'' and ``right'' edges are wrapped, the
``bottom'' edge has a constant value up to a depth of two elements, and
the corner cases, $(-1, -1)$ and $(+1, -1)$, reflect the
corresponding corner elements inside the grid:
\begin{Verbatim}
[boundary| Double  (*i, -1) g -> g!!!(i, 0)                       -- top
                   (-1, *j) g -> g!!!(fst (size g) - 1, j)        -- left
                   (+1, *j) g -> g!!!(0, j)                       -- right
                   from (-1, +1) to (+1, +2) -> 0.0               -- bottom
                   (-1, -1) g -> g!!!(0, 0)                       -- top corners 
                   (+1, -1) g -> g!!!(fst (size g) - 1, 0) |]
\end{Verbatim}
The \ttt{(!!!)} operator is a safe general indexing operation (i.e. 
out-of-bounds access raises an exception) that is only
available within a \ttt{boundary} expression. Bounds checks for 
\ttt{(!!!)} do not have a significant effect on program performance as
such a boundary is a relatively small percentage of the overall grid
size and is only calculated once per application of a stencil
function to update the boundary values from the updated grid values. Note, a
boundary without any parameterised regions is constructed just once
overall as its values do not, and cannot, depend on a grid's inner values.

The boundary definition of a grid must provide sufficient
cover for a stencil function such that the stencil function does not
access an element which is outside of the grid or boundary
region. Since grid patterns and boundary definitions are
static this safety property can be checked statically. Any
program without appropriate boundaries for a stencil application is
rejected.

\section{Statically Enforcing Safe Indexing via Types}
\label{types}


Ypnos provides efficiency and correctness guarantees by 
enforcing safe indexing statically via typing. Thus, well-typed Ypnos
programs are safe, and bounds-check free.

Type-level information
in Ypnos is constructed, composed, and manipulated using various
type-system features of Haskell and the
Glasgow Haskell Compiler. \emph{Generalised
  algebraic data types} (GADTs) \cite{jones2004wobbly,
  peyton2006simple} are used as a form of lightweight dependent-typing,
propagating information from 
data-level to type-level via the constructors of a data
type. \emph{Type families} \cite{associated-type-synonyms} are used as
simple type-level functions to manipulate/compose type-level
information. \emph{Type classes} \cite{hall1996type} are used as
predicates and relations on types. Appendix \ref{haskell-types} provides a brief
introduction to these type-system features for the unfamiliar reader.

We introduce here the core aspects of the Ypnos implementation that
are most relevant to enforcing safe indexing. This section also
explains, mostly by example, the desugaring of Ypnos syntax into Haskell.
We begin with an informal high-level overview of how safety is
enforced before delving into details and definitions.
So far, Ypnos programs have been typeset in \ttt{monospace} to
present a clear view of the concrete syntax. The Haskell
code shown here is typeset more elaborately to ease reading.

\subsection{Safety Overview}


The key to enforcing safety in Ypnos is the encoding of relative
indices at the type-level. Type-level relative indices are used in the
types of stencil functions and grids in two different ways:
respectively, to encode the relative indexing pattern of a stencil function
and to encode the relative position of boundary regions with
respect to the edge of a grid.


The type of a grid is parameterised by type-level information about the
grid's boundary regions. The grid data type takes four type parameters: 
\begin{hscode}\SaveRestoreHook
\column{B}{@{}>{\hspre}l<{\hspost}@{}}%
\column{E}{@{}>{\hspre}l<{\hspost}@{}}%
\>[B]{}\Conid{Grid}\;\Varid{d}\;\Varid{b}\;\Varid{dyn}\;\Varid{a}{}\<[E]%
\ColumnHook
\end{hscode}\resethooks
where \emph{d} is a dimensionality type, \emph{b} is a type-level
list of the relative indices (with respect to the grid's edge) of the
grid's boundary regions, and \emph{a} is the element type of the grid
(\emph{dyn} will be discussed later).

Stencil functions include in their type the relative indices accessed
by the function, as described by a grid pattern. Grid patterns are
desugared via the \texttt{fun} macro into relative indexing operations
on a grid, with type approximately of the form:
\begin{hscode}\SaveRestoreHook
\column{B}{@{}>{\hspre}l<{\hspost}@{}}%
\column{11}{@{}>{\hspre}l<{\hspost}@{}}%
\column{E}{@{}>{\hspre}l<{\hspost}@{}}%
\>[B]{}\Varid{index}\mathbin{::}{}\<[11]%
\>[11]{}\Conid{Safe}\;\Varid{i}\;\Varid{b}\Rightarrow \Varid{i}\to \Conid{Grid}\;\Varid{d}\;\Varid{b}\;\Varid{dyn}\;\Varid{a}\to \Varid{a}{}\<[E]%
\ColumnHook
\end{hscode}\resethooks
where \ensuremath{\Varid{index}} takes a relative index of type \ensuremath{\Varid{i}} and a grid of element type \ensuremath{\Varid{a}},
returning a single value of type \ensuremath{\Varid{a}}. The \ensuremath{\Conid{Safe}\;\Varid{i}\;\Varid{b}} constraint enforces safety by requiring that there are sufficient
boundary regions described by \ensuremath{\Varid{b}} such that a relative
index \ensuremath{\Varid{i}} accessed from anywhere within the grid's extent has a defined value.

The boundary information that parameterises a grid type (type
parameter \ensuremath{\Varid{b}} above) is generated from a boundary description used in
constructing a grid. 
The \texttt{boundary} macro desugars a boundary description into a
special data structure which some grid constructors take as a parameter.

The rest of this section is structured as follows: Section
\ref{dimensions} introduces \emph{dimensionality} types, which
determine the index type of a grid. Section
\ref{indices} introduces absolute index types
and the important relative-index type representation. Section
\ref{boundaries} discusses boundary definitions and boundary types,
which leads into Section \ref{grids} on grid constructors where 
boundary information propagates to grid types. Section
\ref{indexing-safety} brings together the different type-level
information, showing how \ensuremath{\Conid{Safe}} is defined to match relative indices
to grid boundaries. Section \ref{stencil} completes by showing
application of stencil functions to grids.

\subsection{Dimensions}
\label{dimensions}

As described above, the \ensuremath{\Conid{Grid}} data type has a type parameter representing
the \emph{dimensionality} of a grid. In the example of
Section \ref{ypnos}, one of the parameters to the constructor
\texttt{listGrid} was a dimensionality term: \ttt{Dim X
  :* Dim Y} specifying that the grid is two-dimensional with named
dimensions \ttt{X} and \ttt{Y}.

Dimension terms are defined by the following GADT, with a
constructor \ensuremath{\Conid{Dim}} for single dimensions and a constructor \ensuremath{\,{:}{^{*}}\,} for the
tensor of a single dimension and a dimension term\footnote{This tensor
 operator is theoretically associative but the
 implementation is simplified by defining a right-associating operator.}:
\begin{hscode}\SaveRestoreHook
\column{B}{@{}>{\hspre}l<{\hspost}@{}}%
\column{6}{@{}>{\hspre}l<{\hspost}@{}}%
\column{E}{@{}>{\hspre}l<{\hspost}@{}}%
\>[B]{}\mathbf{data}\;\Conid{Dim}\;\Varid{d}{}\<[E]%
\\
\>[B]{}\mathbf{data}\;(\,{:}{^{*}}\,)\;\Varid{d}\;\Varid{d'}{}\<[E]%
\\[\blanklineskip]%
\>[B]{}\mathbf{data}\;\Conid{Dimensionality}\;\Varid{d}\;\mathbf{where}{}\<[E]%
\\
\>[B]{}\hsindent{6}{}\<[6]%
\>[6]{}\Conid{Dim}\mathbin{::}\Conid{DimIdentifier}\;\Varid{d}\Rightarrow \Varid{d}\to \Conid{Dimensionality}\;(\Conid{Dim}\;\Varid{d}){}\<[E]%
\\
\>[B]{}\hsindent{6}{}\<[6]%
\>[6]{}(\,{:}{^{*}}\,)\mathbin{::}\Conid{Dimensionality}\;(\Conid{Dim}\;\Varid{d})\to \Conid{Dimensionality}\;\Varid{d'}\to \Conid{Dimensionality}\;(\Conid{Dim}\;\Varid{d}\,{:}{^{*}}\,\Varid{d'}){}\<[E]%
\ColumnHook
\end{hscode}\resethooks
In the first two lines, the data types \ensuremath{\Conid{Dim}} and \ensuremath{(\,{:}{^{*}}\,)} are \emph{empty
  declarations} defining type constructors without data
constructors. The data constructors of \ensuremath{\Conid{Dimensionality}} reuse these
names\footnote{The data constructors could have been given different
  names to their
type representatives, e.g.~\ensuremath{\Conid{DimIdent}\mathbin{::}\Varid{d}\to \Conid{Dimensionality}\;(\Conid{Dim}\;\Varid{d})},
but we find the correspondence between data constructors and type
representatives more instructive.}
 and use these empty types in the result type's parameter as type
representatives for the data constructors.
The types of dimensionality terms are \emph{singleton types}, i.e. types with just one
 inhabitant, thus their type uniquely determines their value. For
 example, the term \ensuremath{\Conid{Dim}\;\Conid{X}\,{:}{^{*}}\,\Conid{Dim}\;\Conid{Y}} has type \ensuremath{\Conid{Dimensionality}\;(\Conid{Dim}\;\Conid{X}\,{:}{^{*}}\,\Conid{Dim}\;\Conid{Y})}.

The \ensuremath{\Conid{DimIdentifier}} class (which constraints the \ensuremath{\Conid{Dim}} constructor) 
is that of valid dimension identifier
types. Valid dimension identifiers are declared in Ypnos by
a \ttt{[dimension| ... |]} macro, which expands a list of dimension
identifiers into singleton data types, 
representing dimension identifiers, and instances of \ensuremath{\Conid{DimIdentifier}} e.g.
\vspace{-0.7em}
\begin{center}
{\small{\ttt{[dimension| X, Y |]}}} $\quad\leadsto$ \hspace{-1.5em}
\begin{minipage}{0.4\linewidth}
\begin{hscode}\SaveRestoreHook
\column{B}{@{}>{\hspre}l<{\hspost}@{}}%
\column{E}{@{}>{\hspre}l<{\hspost}@{}}%
\>[B]{}\mathbf{data}\;\Conid{X}\mathrel{=}\Conid{X}{}\<[E]%
\\
\>[B]{}\mathbf{data}\;\Conid{Y}\mathrel{=}\Conid{Y}{}\<[E]%
\\
\>[B]{}\mathbf{instance}\;\Conid{DimIdentifier}\;\Conid{X}{}\<[E]%
\\
\>[B]{}\mathbf{instance}\;\Conid{DimIdentifier}\;\Conid{Y}{}\<[E]%
\ColumnHook
\end{hscode}\resethooks
\end{minipage}
\end{center}
\vspace{-2em}

\subsection{Absolute and Relative Indices}
\label{indices}

Internally, grids are indexed by tuples of \ensuremath{\Conid{Int}} values. Given the
dimensionality of a grid the type of \emph{absolute indices} is calculated by
the \ensuremath{\Conid{Index}} type family with instances (of which we show the
first three):
\begin{hscode}\SaveRestoreHook
\column{B}{@{}>{\hspre}l<{\hspost}@{}}%
\column{56}{@{}>{\hspre}c<{\hspost}@{}}%
\column{56E}{@{}l@{}}%
\column{59}{@{}>{\hspre}l<{\hspost}@{}}%
\column{E}{@{}>{\hspre}l<{\hspost}@{}}%
\>[B]{}\mathbf{type}\;\textbf{family}\;\Conid{Index}\;\Varid{t}{}\<[E]%
\\
\>[B]{}\mathbf{type}\;\mathbf{instance}\;\Conid{Index}\;(\Conid{Dim}\;\Varid{x}){}\<[56]%
\>[56]{}\mathrel{=}{}\<[56E]%
\>[59]{}\Conid{Int}{}\<[E]%
\\
\>[B]{}\mathbf{type}\;\mathbf{instance}\;\Conid{Index}\;((\Conid{Dim}\;\Varid{x})\,{:}{^{*}}\,(\Conid{Dim}\;\Varid{y})){}\<[56]%
\>[56]{}\mathrel{=}{}\<[56E]%
\>[59]{}(\Conid{Int},\Conid{Int}){}\<[E]%
\\
\>[B]{}\mathbf{type}\;\mathbf{instance}\;\Conid{Index}\;((\Conid{Dim}\;\Varid{x})\,{:}{^{*}}\,((\Conid{Dim}\;\Varid{y})\,{:}{^{*}}\,(\Conid{Dim}\;\Varid{z}))){}\<[56]%
\>[56]{}\mathrel{=}{}\<[56E]%
\>[59]{}(\Conid{Int},\Conid{Int},\Conid{Int}){}\<[E]%
\ColumnHook
\end{hscode}\resethooks
Unfortunately, tuples in Haskell are not inductively defined thus a fully
general implementation of Ypnos must have an instance of \ensuremath{\Conid{Index}} for every
possible arity of dimensionality i.e. an infinite number.
In practice, many stencil computations do not require more
than four dimensions. We discuss this issue further in Section
\ref{inductive-tuples} and in the rest of the paper give 
tuple-related definitions up to two or three dimensions.


As mentioned, \emph{relative indices} have a type-level encoding which
is key to enforcing safety. Relative indices are given by tuples of a 
data type encoding of integers, \ensuremath{\Conid{IntT}}, providing a type-level
representation of integers.
\ensuremath{\Conid{IntT}} is defined in terms of a data type of inductively
defined natural numbers, \ensuremath{\Conid{Nat}}, for which the result type of each data
constructor provides a type-level representation of the natural number
constructed, as defined by the following GADT:
\begin{hscode}\SaveRestoreHook
\column{B}{@{}>{\hspre}l<{\hspost}@{}}%
\column{4}{@{}>{\hspre}l<{\hspost}@{}}%
\column{E}{@{}>{\hspre}l<{\hspost}@{}}%
\>[B]{}\mathbf{data}\;\Conid{Z}{}\<[E]%
\\
\>[B]{}\mathbf{data}\;\Conid{S}\;\Varid{n}{}\<[E]%
\\[\blanklineskip]%
\>[B]{}\mathbf{data}\;\Conid{Nat}\;\Varid{n}\;\mathbf{where}{}\<[E]%
\\
\>[B]{}\hsindent{4}{}\<[4]%
\>[4]{}\Conid{Z}\mathbin{::}\Conid{Nat}\;\Conid{Z}{}\<[E]%
\\
\>[B]{}\hsindent{4}{}\<[4]%
\>[4]{}\Conid{S}\mathbin{::}\Conid{Nat}\;\Varid{n}\to \Conid{Nat}\;(\Conid{S}\;\Varid{n}){}\<[E]%
\ColumnHook
\end{hscode}\resethooks
Integers are defined by \ensuremath{\Conid{IntT}} in terms of \ensuremath{\Conid{Nat}} as either a
negative or positive natural number: 
\begin{hscode}\SaveRestoreHook
\column{B}{@{}>{\hspre}l<{\hspost}@{}}%
\column{5}{@{}>{\hspre}l<{\hspost}@{}}%
\column{E}{@{}>{\hspre}l<{\hspost}@{}}%
\>[B]{}\mathbf{data}\;\Conid{Neg}\;\Varid{n}{}\<[E]%
\\
\>[B]{}\mathbf{data}\;\Conid{Pos}\;\Varid{n}{}\<[E]%
\\[\blanklineskip]%
\>[B]{}\mathbf{data}\;\Conid{IntT}\;\Varid{n}\;\mathbf{where}{}\<[E]%
\\
\>[B]{}\hsindent{5}{}\<[5]%
\>[5]{}\Conid{Neg}\mathbin{::}\Conid{Nat}\;(\Conid{S}\;\Varid{n})\to \Conid{IntT}\;(\Conid{Neg}\;(\Conid{S}\;\Varid{n})){}\<[E]%
\\
\>[B]{}\hsindent{5}{}\<[5]%
\>[5]{}\Conid{Pos}\mathbin{::}\Conid{Nat}\;\Varid{n}\to \Conid{IntT}\;(\Conid{Pos}\;\Varid{n}){}\<[E]%
\ColumnHook
\end{hscode}\resethooks
Note that zero has a unique representation: \ensuremath{\Conid{Pos}\;\Conid{Z}}. The expression \ensuremath{\Conid{Neg}\;\Conid{Z}}
is not well-typed as the \ensuremath{\Conid{Neg}} data constructor expects a natural number of at
least one i.e. of type \ensuremath{\Conid{Nat}\;(\Conid{S}\;\Varid{n})}.


The \ensuremath{\Conid{Grid}} data type is parameterised by a type-level
list of the relative positions of boundary regions. An example type-level
list construction is the following GADT of heterogeneous lists:
\begin{hscode}\SaveRestoreHook
\column{B}{@{}>{\hspre}l<{\hspost}@{}}%
\column{4}{@{}>{\hspre}l<{\hspost}@{}}%
\column{E}{@{}>{\hspre}l<{\hspost}@{}}%
\>[B]{}\mathbf{data}\;\Conid{Nil}{}\<[E]%
\\
\>[B]{}\mathbf{data}\;\Conid{Cons}\;\Varid{x}\;\Varid{xs}{}\<[E]%
\\[\blanklineskip]%
\>[B]{}\mathbf{data}\;\Conid{List}\;\Varid{t}\;\mathbf{where}{}\<[E]%
\\
\>[B]{}\hsindent{4}{}\<[4]%
\>[4]{}\Conid{Nil}\mathbin{::}\Conid{List}\;\Conid{Nil}{}\<[E]%
\\
\>[B]{}\hsindent{4}{}\<[4]%
\>[4]{}\Conid{Cons}\mathbin{::}\Varid{x}\to \Conid{List}\;\Varid{xs}\to \Conid{List}\;(\Conid{Cons}\;\Varid{x}\;\Varid{xs}){}\<[E]%
\ColumnHook
\end{hscode}\resethooks
The \ensuremath{\Conid{List}} type thus encodes the types of its elements in its type
parameter \ensuremath{\Varid{t}} as a type-level list e.g.
\begin{equation*}
\ensuremath{(\Conid{Cons}\;\mathrm{1}\;(\Conid{Cons}\;\text{\tt \char34 hello\char34}\;(\Conid{Cons}\;\text{\tt 'a'}\;\Conid{Nil})))\mathbin{::}(\Conid{List}\;(\Conid{Cons}\;\Conid{Int}\;(\Conid{Cons}\;\Conid{String}\;(\Conid{Cons}\;\Conid{Char}\;\Conid{Nil}))))}
\end{equation*}

The type-level representation of a grid's boundary regions uses
the same type-level list representation, but the list contains relative
indices describing the position of boundary regions relative to
the grid's edge. For example, a one-dimensional grid with boundary
regions of \ttt{-1} and \ttt{+1} has a type like:
\begin{hscode}\SaveRestoreHook
\column{B}{@{}>{\hspre}l<{\hspost}@{}}%
\column{E}{@{}>{\hspre}l<{\hspost}@{}}%
\>[B]{}\Conid{Grid}\;(\Conid{Dim}\;\Varid{d})\;(\Conid{Cons}\;(\Conid{IntT}\;(\Conid{Neg}\;(\Conid{S}\;\Conid{Z})))\;(\Conid{Cons}\;(\Conid{IntT}\;(\Conid{Pos}\;(\Conid{S}\;\Conid{Z})))\;\Conid{Nil}))\;\Varid{dyn}\;\Varid{a}{}\<[E]%
\ColumnHook
\end{hscode}\resethooks
\vspace{-2em}


%
%
%
\subsection{Boundaries}
\label{boundaries}

A relative index cannot be accessed safely from everywhere within a
grid's extent unless the grid has appropriate
boundary values, provided by a boundary definition.
Ypnos boundary definitions are desugared by the \ttt{boundary} macro
into a data structure describing the boundary which encodes in its type
the positions of the boundary elements defined, relative to the edge of the
grid. This type-level boundary information is propagated
to a grid's type (the \ensuremath{\Varid{b}} type parameter in the \ensuremath{\Conid{Grid}} type). The
\ensuremath{\Conid{Safe}} constraint in the type of relative indexing operations
uses this boundary information to ascertain if a grid has adequate
boundaries for the relative index to be safe when accessed from anywhere
within the grid's extent.

Section \ref{boundaries-intro} introduced Ypnos's syntax for describing
boundaries, with the grammar:
\begin{align*}
B \; ::= & \;\; \texttt{from} \;\; I_1 \;\; \texttt{to} \;\; I_2 \;\;
 \texttt{->} \;\; e \;\; | \;\; I \;\; \texttt{->} \;\; e
 \;\; | \;\; I \;\; g \;\; \texttt{->} \;\; e \;\; \\
I \; ::= & \;\; P \;\; | \;\; \texttt{($P$, $P$)} \;\; | \;\; \texttt{($P$,
  $P$, $P$)} \;\; | \;\; \texttt{($P$, $P$, $P$, $P$)} \;\; | \;\;
\ldots \\
P \; ::= & \;\; \texttt{-}n \;\; | \;\; \texttt{+}n \;\; | \;\; \texttt{*}v 
\end{align*}
The syntax, $I \;\; g \;\;
\texttt{->} \;\; e$, describes boundary
values that are dependent on the contents of a grid \ttt{g}.
Such a boundary is described as \emph{dynamic}.
After the application of a stencil function to a grid
with a dynamic boundary, the boundary values must be recomputed so
that its values are consistent with the grid's (possibly changed) inner
elements. Conversely, a non-parameterised boundary description is
\emph{static}, since it does not depend on the values of the grid and
thus does not need recomputing. Since the \ttt{from...to...} syntax is
desugared into a number of specific forms \ttt{...->...}, as described
in Section \ref{boundaries-intro}, boundary definitions consist of a
number of \ensuremath{\Conid{B}} terms of either the second or third syntactic form.

Boundary definitions are desugared by the \ttt{boundary} macro into a
\ensuremath{\Conid{BoundaryList}} structure which comprises a list of \ensuremath{\Conid{BoundaryFun}} values.
Each \ensuremath{\Conid{B}} term in a boundary definition is desugared into a function
mapping one or more boundary indices, determined by the region
descriptor, to a value. From such functions a \ensuremath{\Conid{BoundaryFun}} value is
constructed 
%
%
which encodes information in its type about the boundary region it defines.
%
%
\ensuremath{\Conid{BoundaryFun}} has two constructors for
static and dynamic regions:
\begin{hscode}\SaveRestoreHook
\column{B}{@{}>{\hspre}l<{\hspost}@{}}%
\column{6}{@{}>{\hspre}l<{\hspost}@{}}%
\column{15}{@{}>{\hspre}c<{\hspost}@{}}%
\column{15E}{@{}l@{}}%
\column{19}{@{}>{\hspre}l<{\hspost}@{}}%
\column{54}{@{}>{\hspre}c<{\hspost}@{}}%
\column{54E}{@{}l@{}}%
\column{58}{@{}>{\hspre}l<{\hspost}@{}}%
\column{E}{@{}>{\hspre}l<{\hspost}@{}}%
\>[B]{}\mathbf{data}\;\Conid{Static}{}\<[E]%
\\
\>[B]{}\mathbf{data}\;\Conid{Dynamic}{}\<[E]%
\\[\blanklineskip]%
\>[B]{}\mathbf{data}\;\Conid{BoundaryFun}\;\Varid{d}\;\Varid{ix}\;\Varid{a}\;\Varid{dyn}\;\mathbf{where}{}\<[E]%
\\
\>[B]{}\hsindent{6}{}\<[6]%
\>[6]{}\Conid{Static}{}\<[15]%
\>[15]{}\mathbin{::}{}\<[15E]%
\>[19]{}(\Varid{ix}\to \Varid{a}){}\<[54]%
\>[54]{}\to {}\<[54E]%
\>[58]{}\Conid{BoundaryFun}\;\Varid{d}\;\Varid{ix}\;\Varid{a}\;\Conid{Static}{}\<[E]%
\\
\>[B]{}\hsindent{6}{}\<[6]%
\>[6]{}\Conid{Dynamic}{}\<[15]%
\>[15]{}\mathbin{::}{}\<[15E]%
\>[19]{}((\Varid{ix},\Conid{Grid}\;\Varid{d}\;\Conid{Nil}\;\Conid{Static}\;\Varid{a})\to \Varid{a}){}\<[54]%
\>[54]{}\to {}\<[54E]%
\>[58]{}\Conid{BoundaryFun}\;\Varid{d}\;\Varid{ix}\;\Varid{a}\;\Conid{Dynamic}{}\<[E]%
\ColumnHook
\end{hscode}\resethooks
Both the \ensuremath{\Conid{Static}} and \ensuremath{\Conid{Dynamic}} constructors of \ensuremath{\Conid{BoundaryFun}} have a
single parameter: a function mapping one or more boundary indices of
type \ensuremath{\Varid{ix}} to a value of type \ensuremath{\Varid{a}}. In the case of a \ensuremath{\Conid{Dynamic}} boundary
function, this index is paired with a grid of element type
\ensuremath{\Varid{a}}. \ensuremath{\Conid{BoundaryFun}} has four type parameters: \ensuremath{\Varid{d}} the
grid's dimensionality, \ensuremath{\Varid{ix}} the boundary indices type, \ensuremath{\Varid{a}} the
element type, and \ensuremath{\Varid{dyn}} denoting whether the boundary is static or
dynamic. 

The region descriptor of a boundary (non-terminal $I$ in the
grammar) determines the type of boundary indices. The components of
a boundary index\footnote{The components of an index are the
  individual elements of the index's tuple for each dimension.}
are either relative or absolute:
%
%
$\texttt{-}n$ and $\texttt{+}n$ are relative
indices, relative to the lower or upper bound respectively of the
dimension's extent, and $\texttt{*}v$ is an absolute index (i.e. an \ensuremath{\Conid{Int}} value) within the
the dimension's extent. 
The region descriptor syntax is desugared into pattern
matches on relative/absolute indices in the following way:
\vspace{-0.8em}
\begin{center}
\begin{minipage}{0.4\linewidth}
\begin{center}
\begin{tabular}{l|l|l}
Syntax & Pattern & Type \\ \hline
\texttt{-n} & \ensuremath{\Conid{Neg}} $\interp{n}$ & $\ensuremath{\Conid{IntT}} \; (\ensuremath{\Conid{Neg}} \; \interp{n})$ \\
\texttt{+n} & \ensuremath{\Conid{Pos}} $\interp{n}$ & $\ensuremath{\Conid{IntT}} \; (\ensuremath{\Conid{Pos}} \; \interp{n})$ \\
\texttt{*v} & v & \ensuremath{\Conid{Int}}
\end{tabular}
\end{center}
\end{minipage}
\begin{minipage}{0.4\linewidth}
\begin{align*}
\text{where} \;\quad \interp{0} & \leadsto \; Z \\
\interp{n} & \leadsto \; S \; \interp{n-1}
\end{align*}
\hspace{3em} at both the data- and type-level.
\vspace{1.2em}
\end{minipage}
\end{center}

\begin{figure}[h]
\centering
\vspace{0.4em}
\hspace{0.15em}
\begin{minipage}{0.2\linewidth}
\scalebox{0.32}{\includegraphics{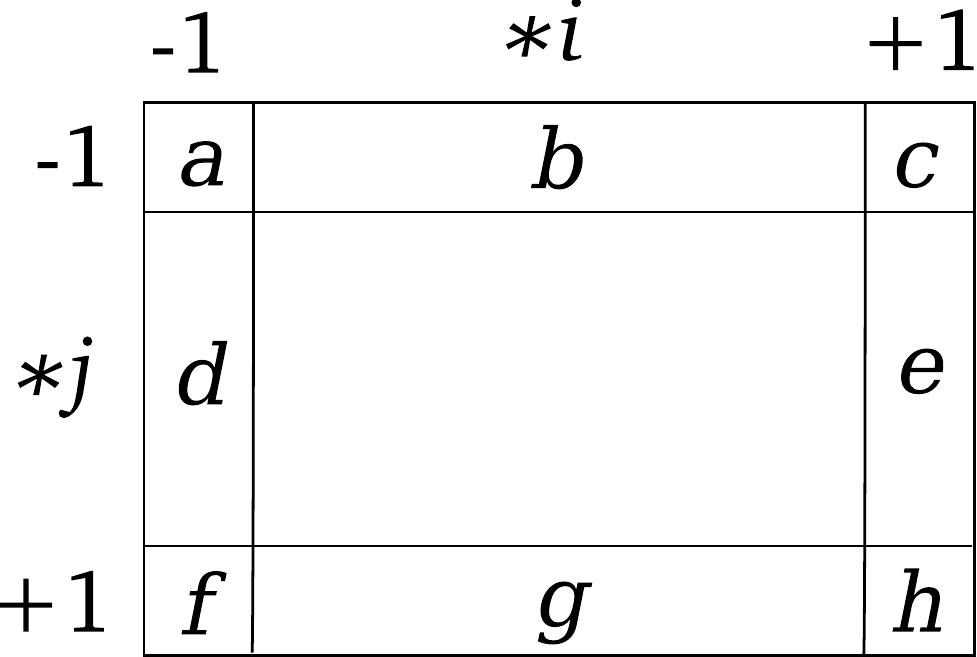}}
\end{minipage}
\hspace{0.25em}
{\scalebox{0.935}{
\begin{minipage}{0.75\linewidth}
\vspace{-1em}
\begin{align*}
\begin{array}{ll}
a = \ensuremath{(\Conid{IntT}\;(\Conid{Neg}\;(\Conid{S}\;\Conid{Z})),\Conid{IntT}\;(\Conid{Neg}\;(\Conid{S}\;\Conid{Z})))} & e = \ensuremath{(\Conid{IntT}\;(\Conid{Pos}\;(\Conid{S}\;\Conid{Z})),\Conid{Int})} \\
b = \ensuremath{(\Conid{Int},\Conid{IntT}\;(\Conid{Neg}\;(\Conid{S}\;\Conid{Z})))} & f = \ensuremath{(\Conid{IntT}\;(\Conid{Neg}\;(\Conid{S}\;\Conid{Z})),\Conid{IntT}\;(\Conid{Pos}\;(\Conid{S}\;\Conid{Z})))} \\
c = \ensuremath{(\Conid{IntT}\;(\Conid{Pos}\;(\Conid{S}\;\Conid{Z})),\Conid{IntT}\;(\Conid{Neg}\;(\Conid{S}\;\Conid{Z})))} & g = \ensuremath{(\Conid{Int},\Conid{IntT}\;(\Conid{Pos}\;(\Conid{S}\;\Conid{Z})))} \\
d = \ensuremath{(\Conid{IntT}\;(\Conid{Neg}\;(\Conid{S}\;\Conid{Z})),\Conid{Int})} &  h = \ensuremath{(\Conid{IntT}\;(\Conid{Pos}\;(\Conid{S}\;\Conid{Z})),\Conid{IntT}\;(\Conid{Pos}\;(\Conid{S}\;\Conid{Z})))}
\end{array}
\end{align*}
\end{minipage}}}
\vspace{1em}
\hrule
\caption{Boundary indices types for a two-dimensional grid with one-element boundary}
\label{boundary-regions-t}
\end{figure}

\myfigref{boundary-regions-t} provides an example of
the types of the boundary indices for the boundary regions of a
two-dimensional grid with a one-element deep boundary.
The following shows the desugaring of two example boundary
definitions into \ensuremath{\Conid{BoundaryFun}} values, with explicit type signatures:
\begin{align*}
& \interp{\ttt{(-1, +1) -> 0.0}} \\
& \leadsto \; \ensuremath{(\Conid{Static}\;(\lambda (\Conid{Neg}\;(\Conid{S}\;\Conid{Z}),\Conid{Pos}\;(\Conid{S}\;\Conid{Z}))\to \mathrm{0.0}))} \\
& \quad\quad \ensuremath{\mathbin{::}(\Conid{BoundaryFun}\;(\Conid{Dim}\;\Varid{d}\,{:}{^{*}}\,\Conid{Dim}\;\Varid{d'})\;(\Conid{IntT}\;(\Conid{Neg}\;(\Conid{S}\;\Conid{Z})),\Conid{IntT}\;(\Conid{Pos}\;(\Conid{S}\;\Conid{Z})))\;\Conid{Double}\;\Conid{Static})} \\
& \interp{\ttt{(*i, +2) g -> g!!!(i, 1)}} \\
& \leadsto \; \ensuremath{(\Conid{Dynamic}\;(\lambda ((\Varid{i},\Conid{Pos}\;(\Conid{S}\;(\Conid{S}\;\Conid{Z}))),\Varid{g})\to \Varid{g}\mathbin{!!!}(\Varid{i},\mathrm{1})))} \\ 
& \quad\quad \ensuremath{\mathbin{::}(\Conid{BoundaryFun}\;(\Conid{Dim}\;\Varid{d}\,{:}{^{*}}\,\Conid{Dim}\;\Varid{d'})\;(\Conid{Int},\Conid{IntT}\;(\Conid{Pos}\;(\Conid{S}\;(\Conid{S}\;\Conid{Z}))))\;\Conid{Double}\;\Conid{Dynamic})}
\end{align*}
Functions with GADT parameters require explicit type signatures
\cite{peyton2006simple} thus the \texttt{boundary} macro 
generates such type signatures for the boundary functions it defines,
of which we showed two examples above. The boundary index types and
the type for static or dynamic boundaries can be easily generated
from the syntax of a boundary definition during desugaring. However,
the element type of a grid/boundary cannot be constructed without
performing type inference. To simplify the implementation the user
must provide a monomorphic type for the boundary values at the start
of a boundary definition, as seen in the Laplace example where the
\ensuremath{\Conid{Double}} type is specified.

\ensuremath{\Conid{BoundaryFun}} values are composed via the \ensuremath{\Conid{BoundaryList}} structure which is
similar to the \ensuremath{\Conid{List}} GADT seen earlier but with additional
type-level information, defined:

\begin{minipage}{1\linewidth}
\begin{hscode}\SaveRestoreHook
\column{B}{@{}>{\hspre}l<{\hspost}@{}}%
\column{5}{@{}>{\hspre}l<{\hspost}@{}}%
\column{15}{@{}>{\hspre}l<{\hspost}@{}}%
\column{32}{@{}>{\hspre}l<{\hspost}@{}}%
\column{E}{@{}>{\hspre}l<{\hspost}@{}}%
\>[B]{}\mathbf{data}\;\Conid{BoundaryList}\;\Varid{b}\;\Varid{dyn}\;\Varid{lower}\;\Varid{upper}\;\Varid{d}\;\Varid{a}\;\mathbf{where}{}\<[E]%
\\
\>[B]{}\hsindent{5}{}\<[5]%
\>[5]{}\Conid{NilB}\mathbin{::}\Conid{BoundaryList}\;\Conid{Nil}\;\Conid{Static}\;(\Conid{Origin}\;\Varid{d})\;(\Conid{Origin}\;\Varid{d})\;\Varid{d}\;\Varid{a}{}\<[E]%
\\
\>[B]{}\hsindent{5}{}\<[5]%
\>[5]{}\Conid{ConsB}\mathbin{::}{}\<[15]%
\>[15]{}\Conid{BoundaryFun}\;\Varid{d}\;\Varid{ix}\;\Varid{a}\;\Varid{dyn}{}\<[E]%
\\
\>[15]{}\to \Conid{BoundaryList}\;\Varid{b}\;\Varid{dyn'}\;\Varid{lower}\;\Varid{upper}\;\Varid{d}\;\Varid{a}{}\<[E]%
\\
\>[15]{}\to \Conid{BoundaryList}\;{}\<[32]%
\>[32]{}(\Conid{Cons}\;(\Conid{AbsToRel}\;\Varid{ix})\;\Varid{b})\;(\Conid{Dynamism}\;\Varid{dyn}\;\Varid{dyn'})\;{}\<[E]%
\\
\>[32]{}(\Conid{Lower}\;\Varid{ix}\;\Varid{lower})\;(\Conid{Upper}\;\Varid{ix}\;\Varid{upper})\;\Varid{d}\;\Varid{a}{}\<[E]%
\ColumnHook
\end{hscode}\resethooks
\end{minipage}

The \ensuremath{\Conid{BoundaryList}} structure has six type-parameters encoding
inductively-defined information about the boundary regions via the \ensuremath{\Conid{NilB}}
constructor for the base case and \ensuremath{\Conid{ConsB}} constructor for the
inductive case. The first type-parameter \ensuremath{\Varid{b}} is a type-level list, defined
using the \ensuremath{\Conid{Nil}} and \ensuremath{\Conid{Cons}} type constructors, which encodes the
boundary indices defined by the \ensuremath{\Conid{BoundaryList}}.
In the inductive case, the \ensuremath{\Conid{AbsToRel}} type family is applied to the
boundary indices type \ensuremath{\Varid{ix}} of the \ensuremath{\Conid{BoundaryFun}}. \ensuremath{\Conid{AbsToRel}} converts a
boundary indices type into relative index type. 
Boundary indices have components that are either absolute or relative
indices. \ensuremath{\Conid{AbsToRel}} converts absolute indices or type \ensuremath{\Conid{Int}}, which are
indices within the extent of a dimension, to zero-relative indices
as an index within the extent of a dimension is neither greater
than or less than the extent of a dimension. \ensuremath{\Conid{AbsToRel}} is defined:
%
%
\begin{hscode}\SaveRestoreHook
\column{B}{@{}>{\hspre}l<{\hspost}@{}}%
\column{59}{@{}>{\hspre}l<{\hspost}@{}}%
\column{63}{@{}>{\hspre}l<{\hspost}@{}}%
\column{E}{@{}>{\hspre}l<{\hspost}@{}}%
\>[B]{}\mathbf{type}\;\textbf{family}\;\Conid{AbsToRel}\;\Varid{t}{}\<[E]%
\\
\>[B]{}\mathbf{type}\;\mathbf{instance}\;\Conid{AbsToRel}\;\Conid{Int}\mathrel{=}\Conid{IntT}\;(\Conid{Pos}\;\Conid{Z})\;{}\<[59]%
\>[59]{}\,{}\<[63]%
\>[63]{}\mbox{\onelinecomment  \ensuremath{\Conid{Int}} to zero (\ensuremath{\Conid{IntT}\;(\Conid{Pos}\;\Conid{Z})})}{}\<[E]%
\\
\>[B]{}\mathbf{type}\;\mathbf{instance}\;\Conid{AbsToRel}\;(\Conid{IntT}\;\Varid{n})\mathrel{=}\Conid{IntT}\;\Varid{n}\;{}\<[59]%
\>[59]{}\,{}\<[63]%
\>[63]{}\mbox{\onelinecomment  \ensuremath{\Conid{IntT}} to \ensuremath{\Conid{IntT}}}{}\<[E]%
\\
\>[B]{}\mathbf{type}\;\mathbf{instance}\;\Conid{AbsToRel}\;(\Varid{a},\Varid{b})\mathrel{=}(\Conid{AbsToRel}\;\Varid{a},\Conid{AbsToRel}\;\Varid{b})\;{}\<[59]%
\>[59]{}\,{}\<[63]%
\>[63]{}\mbox{\onelinecomment  Two-dimensional case}{}\<[E]%
\ColumnHook
\end{hscode}\resethooks
%
The type-level list of boundary indices is thus a list of relative
indices with respect to the grid's lower or upper extent, where a
zero-relative index component means any index within the grid's
extent. This representation
is useful for matching relative indices from indexing operations 
with the boundary information of a grid, to which we return in
Section \ref{indexing-safety}).

The remaining parameters to \ensuremath{\Conid{BoundaryList}} encode other information
which is used by the implementation but is not involved in enforcing
safety. Briefly, \ensuremath{\Varid{dyn}} describes whether all boundary sub-definitions are
    static or if at least one is dynamic. The base case is that all
    are \ensuremath{\Conid{Static}}. The inductive
    case applies the \ensuremath{\Conid{Dynamism}} type family, which is monotonic (in the
    two element domain $\{\ensuremath{\Conid{Static}}, \ensuremath{\Conid{Dynamic}}\}$), calculating the
    dynamic property for the entire boundary:
\begin{hscode}\SaveRestoreHook
\column{B}{@{}>{\hspre}l<{\hspost}@{}}%
\column{E}{@{}>{\hspre}l<{\hspost}@{}}%
\>[B]{}\mathbf{type}\;\textbf{family}\;\Conid{Dynamism}\;\Varid{t}\;\Varid{t'}{}\<[E]%
\\
\>[B]{}\mathbf{type}\;\mathbf{instance}\;\Conid{Dynamism}\;\Conid{Static}\;\Conid{Static}\mathrel{=}\Conid{Static}{}\<[E]%
\\
\>[B]{}\mathbf{type}\;\mathbf{instance}\;\Conid{Dynamism}\;\Conid{Static}\;\Conid{Dynamic}\mathrel{=}\Conid{Dynamic}{}\<[E]%
\\
\>[B]{}\mathbf{type}\;\mathbf{instance}\;\Conid{Dynamism}\;\Conid{Dynamic}\;\Conid{Static}\mathrel{=}\Conid{Dynamic}{}\<[E]%
\\
\>[B]{}\mathbf{type}\;\mathbf{instance}\;\Conid{Dynamism}\;\Conid{Dynamic}\;\Conid{Dynamic}\mathrel{=}\Conid{Dynamic}{}\<[E]%
\ColumnHook
\end{hscode}\resethooks
The \ensuremath{\Varid{lower}} and \ensuremath{\Varid{upper}} parameters encode the lower and upper bounds
of the boundary regions, relative to the lower and upper bounds of the
grid's extent, which is used by the implementation
to construct the data array for a grid. The base case for
both is the \ensuremath{\Conid{Origin}} for a particular dimensionality, which is the
zero-relative index e.g. \ensuremath{\Conid{Origin}\;((\Conid{Dim}\;\Varid{d})\,{:}{^{*}}\,(\Conid{Dim}\;\Varid{d'}))\mathrel{=}(\Conid{IntT}\;(\Conid{Pos}\;\Conid{Z}),\Conid{IntT}\;(\Conid{Pos}\;\Conid{Z}))}.
The inductive cases are calculated by the \ensuremath{\Conid{Lower}} and \ensuremath{\Conid{Upper}} type
families which compute the lower and upper bound of a pair of boundary
indices, defined as the minimum and maximum of each
component of an index.

Lastly, the \ensuremath{\Varid{d}} parameter of \ensuremath{\Conid{BoundaryList}} is the dimensionality of the grid for which
this is the boundary, and \ensuremath{\Varid{a}} is the element type of the grid/boundary.

In the Laplace example of Section \ref{ypnos}, the type of the boundary
definition, \ttt{laplaceBoundary}, is:
\begin{hscode}\SaveRestoreHook
\column{B}{@{}>{\hspre}l<{\hspost}@{}}%
\column{15}{@{}>{\hspre}l<{\hspost}@{}}%
\column{28}{@{}>{\hspre}l<{\hspost}@{}}%
\column{38}{@{}>{\hspre}c<{\hspost}@{}}%
\column{38E}{@{}l@{}}%
\column{41}{@{}>{\hspre}l<{\hspost}@{}}%
\column{57}{@{}>{\hspre}l<{\hspost}@{}}%
\column{61}{@{}>{\hspre}l<{\hspost}@{}}%
\column{70}{@{}>{\hspre}l<{\hspost}@{}}%
\column{E}{@{}>{\hspre}l<{\hspost}@{}}%
\>[B]{}\Conid{BoundaryList}\;{}\<[15]%
\>[15]{}(\Conid{Cons}\;(\Conid{IntT}\;(\Conid{Neg}\;(\Conid{S}\;\Conid{Z})),{}\<[41]%
\>[41]{}\Conid{IntT}\;(\Conid{Neg}\;(\Conid{S}\;\Conid{Z})))\;{}\<[61]%
\>[61]{}\,\;\,{}\<[70]%
\>[70]{}\mbox{\onelinecomment  \ttt{(-1, -1)}}{}\<[E]%
\\
\>[15]{}(\Conid{Cons}\;(\Conid{IntT}\;(\Conid{Neg}\;(\Conid{S}\;\Conid{Z})),{}\<[41]%
\>[41]{}\Conid{IntT}\;(\Conid{Pos}\;\Conid{Z}))\;{}\<[61]%
\>[61]{}\,\;\,{}\<[70]%
\>[70]{}\mbox{\onelinecomment  \ttt{(-1, *j)}}{}\<[E]%
\\
\>[15]{}(\Conid{Cons}\;(\Conid{IntT}\;(\Conid{Neg}\;(\Conid{S}\;\Conid{Z})),{}\<[41]%
\>[41]{}\Conid{IntT}\;(\Conid{Pos}\;(\Conid{S}\;\Conid{Z})))\;{}\<[61]%
\>[61]{}\,\;\,{}\<[70]%
\>[70]{}\mbox{\onelinecomment  \ttt{(-1, +1)}}{}\<[E]%
\\
\>[15]{}(\Conid{Cons}\;(\Conid{IntT}\;(\Conid{Pos}\;\Conid{Z}){}\<[38]%
\>[38]{},{}\<[38E]%
\>[41]{}\Conid{IntT}\;(\Conid{Neg}\;(\Conid{S}\;\Conid{Z})))\;{}\<[61]%
\>[61]{}\,\;\,{}\<[70]%
\>[70]{}\mbox{\onelinecomment  \ttt{(*i, -1)}}{}\<[E]%
\\
\>[15]{}(\Conid{Cons}\;(\Conid{IntT}\;(\Conid{Pos}\;\Conid{Z}){}\<[38]%
\>[38]{},{}\<[38E]%
\>[41]{}\Conid{IntT}\;(\Conid{Pos}\;(\Conid{S}\;\Conid{Z})))\;{}\<[61]%
\>[61]{}\,\;\,{}\<[70]%
\>[70]{}\mbox{\onelinecomment  \ttt{(*i, +1)}}{}\<[E]%
\\
\>[15]{}(\Conid{Cons}\;(\Conid{IntT}\;(\Conid{Pos}\;(\Conid{S}\;\Conid{Z})),{}\<[41]%
\>[41]{}\Conid{IntT}\;(\Conid{Neg}\;(\Conid{S}\;\Conid{Z})))\;{}\<[61]%
\>[61]{}\,\;\,{}\<[70]%
\>[70]{}\mbox{\onelinecomment  \ttt{(+1, -1)}}{}\<[E]%
\\
\>[15]{}(\Conid{Cons}\;(\Conid{IntT}\;(\Conid{Pos}\;(\Conid{S}\;\Conid{Z})),{}\<[41]%
\>[41]{}\Conid{IntT}\;(\Conid{Pos}\;\Conid{Z}){}\<[57]%
\>[57]{})\;{}\<[61]%
\>[61]{}\,\;\,{}\<[70]%
\>[70]{}\mbox{\onelinecomment  \ttt{(+1, *j)}}{}\<[E]%
\\
\>[15]{}(\Conid{Cons}\;(\Conid{IntT}\;(\Conid{Pos}\;(\Conid{S}\;\Conid{Z})),{}\<[41]%
\>[41]{}\Conid{IntT}\;(\Conid{Pos}\;(\Conid{S}\;\Conid{Z})))\;{}\<[61]%
\>[61]{}\,\;\,{}\<[70]%
\>[70]{}\mbox{\onelinecomment  \ttt{(+1, +1)}}{}\<[E]%
\\
\>[15]{}\Conid{Nil}))))))))\;{}\<[28]%
\>[28]{}\Conid{Static}\;{}\<[E]%
\\
\>[15]{}(\Conid{IntT}\;(\Conid{Neg}\;(\Conid{S}\;\Conid{Z})),\Conid{IntT}\;(\Conid{Neg}\;(\Conid{S}\;\Conid{Z})))\;(\Conid{IntT}\;(\Conid{Pos}\;(\Conid{S}\;\Conid{Z})),\Conid{IntT}\;(\Conid{Pos}\;(\Conid{S}\;\Conid{Z})))\;{}\<[E]%
\\
\>[15]{}(\Conid{Dim}\;\Conid{X}\,{:}{^{*}}\,\Conid{Dim}\;\Conid{Y})\;\Conid{Double}{}\<[E]%
\ColumnHook
\end{hscode}\resethooks
Fortunately such a type is never constructed by an end-user, although
it is possible for such a type to be seen as an error message (see
Section \ref{errors} for further discussion). Note that for
\ensuremath{\Conid{Dynamism}}, \ensuremath{\Conid{Lower}}, and \ensuremath{\Conid{Upper}} a corresponding
value is not computed, thus these computations occur only at
compile-time.


In the original paper there was some mention of a
boundary structure called a \emph{facets} structure, this
corresponds to the \ensuremath{\Conid{BoundaryList}} structure which has been elaborated
here following recent research.

\subsection{Grids and Grid Constructors}
\label{grids}

A \ensuremath{\Conid{BoundaryList}} structure can be passed to grid constructor to define
the boundary of a grid. The first type parameter of a \ensuremath{\Conid{BoundaryList}}
structure provides type-level information of the boundary it defines
as a list of relative indices of the boundary regions. A grid
constructed with a \ensuremath{\Conid{BoundaryList}} adopts this type-level boundary
information in its type.
%

The \ensuremath{\Conid{Grid}} data type has four type parameters: \ensuremath{\Conid{Grid}\;\Varid{d}\;\Varid{b}\;\Varid{dyn}\;\Varid{a}},
where \emph{d} is a dimensionality type, \emph{b} is the boundary
information, \emph{dyn} describes whether the grid's boundary is
static or dynamic, and \emph{a} is the element type of the grid. 
\ensuremath{\Conid{Grid}} is defined by the following GADT with just one constructor:
\begin{hscode}\SaveRestoreHook
\column{B}{@{}>{\hspre}l<{\hspost}@{}}%
\column{5}{@{}>{\hspre}l<{\hspost}@{}}%
\column{14}{@{}>{\hspre}l<{\hspost}@{}}%
\column{56}{@{}>{\hspre}l<{\hspost}@{}}%
\column{E}{@{}>{\hspre}l<{\hspost}@{}}%
\>[B]{}\mathbf{data}\;\Conid{Grid}\;\Varid{d}\;\Varid{b}\;\Varid{dyn}\;\Varid{a}\;\mathbf{where}{}\<[E]%
\\
\>[B]{}\hsindent{5}{}\<[5]%
\>[5]{}\Conid{Grid}\mathbin{::}{}\<[14]%
\>[14]{}(\Conid{IArray}\;\Conid{UArray}\;\Varid{a}){}\<[E]%
\\
\>[14]{}\Rightarrow (\Conid{UArray}\;(\Conid{Index}\;\Varid{d})\;\Varid{a}){}\<[56]%
\>[56]{}\mbox{\onelinecomment  Array of values}{}\<[E]%
\\
\>[14]{}\to \Conid{Dimensionality}\;\Varid{d}{}\<[56]%
\>[56]{}\mbox{\onelinecomment  Dimensionality term}{}\<[E]%
\\
\>[14]{}\to \Conid{Index}\;\Varid{d}{}\<[56]%
\>[56]{}\mbox{\onelinecomment  Cursor (``current index") }{}\<[E]%
\\
\>[14]{}\to (\Conid{Index}\;\Varid{d},\Conid{Index}\;\Varid{d}){}\<[56]%
\>[56]{}\mbox{\onelinecomment  Lower and upper bounds of grid extent}{}\<[E]%
\\
\>[14]{}\to \Conid{BoundaryList}\;\Varid{b}\;\Varid{dyn}\;\Varid{lower}\;\Varid{upper}\;\Varid{d}\;\Varid{a}{}\<[56]%
\>[56]{}\mbox{\onelinecomment  Boundary definition}{}\<[E]%
\\
\>[14]{}\to \Conid{Grid}\;\Varid{d}\;\Varid{b}\;\Varid{dyn}\;\Varid{a}{}\<[E]%
\ColumnHook
\end{hscode}\resethooks

The type parameters \ensuremath{\Varid{b}} and \ensuremath{\Varid{dyn}} of \ensuremath{\Conid{BoundaryList}}, which
encode boundary regions and dynamism information, are transferred to
the constructed \ensuremath{\Conid{Grid}}.
The \ensuremath{\Conid{IArray}\;\Conid{UArray}\;\Varid{a}} constraint is a consequence of the underlying
implementation of grids as Haskell unboxed array data types.

The \ensuremath{\Conid{Grid}} data constructor is hidden from the end-user. Instead a
number of different constructor functions are provided, the most
common of which are \ensuremath{\Varid{listGrid}} (seen earlier) and \ensuremath{\Varid{grid}}, which have type:
\begin{hscode}\SaveRestoreHook
\column{B}{@{}>{\hspre}l<{\hspost}@{}}%
\column{14}{@{}>{\hspre}l<{\hspost}@{}}%
\column{E}{@{}>{\hspre}l<{\hspost}@{}}%
\>[B]{}\Varid{grid}\mathbin{::}{}\<[14]%
\>[14]{}(\Conid{IArray}\;\Conid{UArray}\;\Varid{a},\Conid{Reifiable}\;\Varid{upper}\;(\Conid{Index}\;\Varid{d}),\Conid{Reifiable}\;\Varid{lower}\;(\Conid{Index}\;\Varid{d}))\Rightarrow {}\<[E]%
\\
\>[14]{}\Conid{Dimensionality}\;\Varid{d}\to \Conid{Index}\;\Varid{d}\to \Conid{Index}\;\Varid{d}\to [\mskip1.5mu (\Conid{Index}\;\Varid{d},\Varid{a})\mskip1.5mu]\to {}\<[E]%
\\
\>[14]{}\Conid{BoundaryList}\;\Varid{b}\;\Varid{dyn}\;\Varid{lower}\;\Varid{upper}\;\Varid{d}\;\Varid{a}\to \Conid{Grid}\;\Varid{d}\;\Varid{b}\;\Varid{dyn}\;\Varid{a}{}\<[E]%
\\[\blanklineskip]%
\>[B]{}\Varid{listGrid}\mathbin{::}{}\<[14]%
\>[14]{}(\Conid{IArray}\;\Conid{UArray}\;\Varid{a},\Conid{Reifiable}\;\Varid{upper}\;(\Conid{Index}\;\Varid{d}),\Conid{Reifiable}\;\Varid{lower}\;(\Conid{Index}\;\Varid{d}))\Rightarrow {}\<[E]%
\\
\>[14]{}\Conid{Dimensionality}\;\Varid{d}\to \Conid{Index}\;\Varid{d}\to \Conid{Index}\;\Varid{d}\to [\mskip1.5mu \Varid{a}\mskip1.5mu]\to {}\<[E]%
\\
\>[14]{}\Conid{BoundaryList}\;\Varid{b}\;\Varid{dyn}\;\Varid{lower}\;\Varid{upper}\;\Varid{d}\;\Varid{a}\to \Conid{Grid}\;\Varid{d}\;\Varid{b}\;\Varid{dyn}\;\Varid{a}{}\<[E]%
\ColumnHook
\end{hscode}\resethooks
The first three parameters of both constructors are a dimensionality
term, a lower-bound index, and an upper-bound index.
The fourth parameter for \ensuremath{\Varid{listGrid}} is a list of elements 
which define the data of the grid, and for \ensuremath{\Varid{grid}} a list of
index-element pairs. The fifth parameter of both is a boundary
definition.

Recall that relative indices are singleton types, thus the type
uniquely determines the value. The \ensuremath{\Conid{Reifiable}} class provides
functions for constructing values from relative index types. In this
case, the \ensuremath{\Varid{upper}} and \ensuremath{\Varid{lower}} bounds of a boundary can be reified as
absolute indices. \ensuremath{\Conid{Reifiable}} is used elsewhere in the implementation
to construct relative index values, or \ensuremath{\Conid{Int}}-valued relative indices,
from a type.
%

Ypnos provides two further constructors
\ensuremath{\Varid{listGridNoBoundary}} and \ensuremath{\Varid{gridNoBoundary}}, not shown here. Both
are similar to \ensuremath{\Varid{listGrid}} and \ensuremath{\Varid{grid}} but are not
parameterised by a \ensuremath{\Conid{BoundaryList}} structure, returning a grid of type \ensuremath{\Conid{Grid}\;\Varid{d}\;\Conid{Nil}\;\Conid{Static}\;\Varid{a}}, where \ensuremath{\Conid{Nil}} boundary information implies no boundary.

\subsection{Grid Patterns, Indexing, and Safety}
\label{indexing-safety}

Grid patterns are desugared into a number of relative indexing
operations which are internally implemented without bounds-checks.
The one- and two-dimension relative indexing operations have type:
\begin{hscode}\SaveRestoreHook
\column{B}{@{}>{\hspre}l<{\hspost}@{}}%
\column{13}{@{}>{\hspre}l<{\hspost}@{}}%
\column{E}{@{}>{\hspre}l<{\hspost}@{}}%
\>[B]{}\Varid{index1D}\mathbin{::}{}\<[13]%
\>[13]{}\Conid{Safe}\;(\Conid{IntT}\;\Varid{n})\;\Varid{b}\Rightarrow {}\<[E]%
\\
\>[13]{}\Conid{IntT}\;\Varid{n}\to \Conid{Int}\to \Conid{Grid}\;(\Conid{Dim}\;\Varid{d})\;\Varid{b}\;\Varid{dyn}\;\Varid{a}\to \Varid{a}{}\<[E]%
\\[\blanklineskip]%
\>[B]{}\Varid{index2D}\mathbin{::}{}\<[13]%
\>[13]{}\Conid{Safe}\;(\Conid{IntT}\;\Varid{n},\Conid{IntT}\;\Varid{n'})\;\Varid{b}\Rightarrow {}\<[E]%
\\
\>[13]{}(\Conid{IntT}\;\Varid{n},\Conid{IntT}\;\Varid{n'})\to (\Conid{Int},\Conid{Int})\to \Conid{Grid}\;(\Conid{Dim}\;\Varid{d}\,{:}{^{*}}\,\Conid{Dim}\;\Varid{d'})\;\Varid{b}\;\Varid{dyn}\;\Varid{a}\to \Varid{a}{}\<[E]%
\ColumnHook
\end{hscode}\resethooks
Note that, in both cases, the first parameter is a relative index and the
second parameter is an absolute index. The \ttt{fun} macro generates
at compile time both an inductively defined relative index, to produce the correct
type constraints, and an \ensuremath{\Conid{Int}}-valued relative index which is used to
perform the actual access. This second parameter is provided so that
an inductive relative index does not have to be (expensively) converted into an
\ensuremath{\Conid{Int}} representation for actual indexing at run-time. 

The \ensuremath{\Conid{Safe}} constraint enforces safety by requiring that a grid's boundary
provides sufficient elements outside of a grid such that a relative
index has a defined value if accessed from anywhere within the grid.
As a first approximation \ensuremath{\Conid{Safe}} is essentially defined as membership of a
relative index to the list of boundary regions for a grid, which are
the relative indices of boundary regions from the edge of the grid.

The \ensuremath{\Conid{InBoundary}} class is used by \ensuremath{\Conid{Safe}} as a predicate to test
whether a relative index matches a relative position of a boundary
region in the list of boundary regions:
\begin{hscode}\SaveRestoreHook
\column{B}{@{}>{\hspre}l<{\hspost}@{}}%
\column{58}{@{}>{\hspre}l<{\hspost}@{}}%
\column{E}{@{}>{\hspre}l<{\hspost}@{}}%
\>[B]{}\mathbf{class}\;\Conid{InBoundary}\;\Varid{i}\;\Varid{ixs}{}\<[E]%
\\
\>[B]{}\mathbf{instance}\;\Conid{InBoundary}\;\Varid{i}\;(\Conid{Cons}\;\Varid{i}\;\Varid{ixs}){}\<[58]%
\>[58]{}\mbox{\onelinecomment  List head matches}{}\<[E]%
\\
\>[B]{}\mathbf{instance}\;\Conid{InBoundary}\;\Varid{i}\;\Varid{ixs}\Rightarrow \Conid{InBoundary}\;\Varid{i}\;(\Conid{Cons}\;\Varid{i'}\;\Varid{ixs}){}\<[58]%
\>[58]{}\mbox{\onelinecomment  List head does not match, recurse}{}\<[E]%
\ColumnHook
\end{hscode}\resethooks
For example, consider a two-dimensional grid with a
one-element deep boundary:
\vspace{0.2em}
\begin{center}
\scalebox{0.30}{\includegraphics{boundary-regions.pdf}}
\end{center}
\vspace{0.2em}
The relative index $(0, +1)$, represented by type \ensuremath{(\Conid{IntT}\;(\Conid{Pos}\;\Conid{Z}),\Conid{IntT}\;(\Conid{Pos}\;(\Conid{S}\;\Conid{Z})))}, requires the \ensuremath{\Varid{g}} boundary to be
safe. The boundary function for \ensuremath{\Varid{g}} pattern
matches on boundary indices of type \ensuremath{(\Conid{Int},\Conid{IntT}\;(\Conid{Pos}\;(\Conid{S}\;\Conid{Z})))}. As \ensuremath{\Conid{AbsToRel}} maps \ensuremath{\Conid{Int}} to a zero-relative index in a
boundary indices type (see Section \ref{boundaries}), the relative
position of \ensuremath{\Varid{g}} is encoded in the type-level
boundary information as: \ensuremath{(\Conid{IntT}\;(\Conid{Pos}\;\Conid{Z}),\Conid{IntT}\;(\Conid{Pos}\;(\Conid{S}\;\Conid{Z})))} i.e. the
same representation as the relative index $(0, +1)$.
%
%
Thus, our first approximation of safety as membership of a relative index to
the list of boundary regions would in this situation be sound. However, 
other relative indices require several boundary regions to be defined.
For example, safety of the relative index $(+1, +1)$ requires the boundaries \ensuremath{\Varid{g}},
\ensuremath{\Varid{h}}, and \ensuremath{\Varid{e}}.

\ensuremath{\Conid{Safe}} is implemented in the following way: for every relative index
we must test the space of possible indices it may access outside of a
grid boundary. For the one-dimensional case \ensuremath{\Conid{Safe}} is defined as such:
\begin{hscode}\SaveRestoreHook
\column{B}{@{}>{\hspre}l<{\hspost}@{}}%
\column{13}{@{}>{\hspre}l<{\hspost}@{}}%
\column{E}{@{}>{\hspre}l<{\hspost}@{}}%
\>[B]{}\mathbf{class}\;\Conid{Safe}\;\Varid{i}\;\Varid{b}{}\<[E]%
\\[\blanklineskip]%
\>[B]{}\mathbf{instance}\;\Conid{Safe}\;(\Conid{IntT}\;(\Conid{Pos}\;\Conid{Z}))\;\Varid{b}{}\<[E]%
\\[\blanklineskip]%
\>[B]{}\mathbf{instance}\;({}\<[13]%
\>[13]{}\Conid{Safe}\;(\Conid{IntT}\;(\Conid{Pred}\;\Varid{n}))\;\Varid{b},\Conid{InBoundary}\;(\Conid{IntT}\;\Varid{n})\;\Varid{b})\Rightarrow \Conid{Safe}\;(\Conid{IntT}\;\Varid{n})\;\Varid{b}{}\<[E]%
\ColumnHook
\end{hscode}\resethooks
The first instance defines that zero-relative indices are always safe.
The second instance defines that a relative index is safe if it is a
member of the boundary regions for a grid and if its predecessor is
also safe. \ensuremath{\Conid{Pred}} is defined:
\begin{hscode}\SaveRestoreHook
\column{B}{@{}>{\hspre}l<{\hspost}@{}}%
\column{51}{@{}>{\hspre}l<{\hspost}@{}}%
\column{E}{@{}>{\hspre}l<{\hspost}@{}}%
\>[B]{}\mathbf{type}\;\textbf{family}\;\Conid{Pred}\;\Varid{n}{}\<[E]%
\\
\>[B]{}\mathbf{type}\;\mathbf{instance}\;\Conid{Pred}\;(\Conid{Neg}\;(\Conid{S}\;(\Conid{S}\;\Varid{n})))\mathrel{=}\Conid{Neg}\;(\Conid{S}\;\Varid{n}){}\<[51]%
\>[51]{}\mbox{\onelinecomment  \emph{Pred} $\,-(n+1) = -n$}{}\<[E]%
\\
\>[B]{}\mathbf{type}\;\mathbf{instance}\;\Conid{Pred}\;(\Conid{Neg}\;(\Conid{S}\;\Conid{Z}))\mathrel{=}\Conid{Pos}\;\Conid{Z}{}\<[51]%
\>[51]{}\mbox{\onelinecomment  \emph{Pred} $\,-1 \quad\quad\; = 0$}{}\<[E]%
\\
\>[B]{}\mathbf{type}\;\mathbf{instance}\;\Conid{Pred}\;(\Conid{Pos}\;\Conid{Z})\mathrel{=}\Conid{Pos}\;\Conid{Z}{}\<[51]%
\>[51]{}\mbox{\onelinecomment  \emph{Pred} $\,0 \quad\quad\quad = 0$}{}\<[E]%
\\
\>[B]{}\mathbf{type}\;\mathbf{instance}\;\Conid{Pred}\;(\Conid{Pos}\;(\Conid{S}\;\Varid{n}))\mathrel{=}\Conid{Pos}\;\Varid{n}{}\<[51]%
\>[51]{}\mbox{\onelinecomment  \emph{Pred} $\,+(n+1) = +n$}{}\<[E]%
\ColumnHook
\end{hscode}\resethooks
Thus, for a relative index of say $(-2)$ there must be a boundary region
for $(-2)$ and $(-1)$, and for a relative index of $(+2)$ there must
be a boundary region $(+2)$ and $(+1)$. \ensuremath{\Conid{Pred}} is thus the predecessor
of a relative index, approaching zero from both the negative and
positive directions.

For the two-dimensional case \ensuremath{\Conid{Safe}} is defined similarly:%
\begin{hscode}\SaveRestoreHook
\column{B}{@{}>{\hspre}l<{\hspost}@{}}%
\column{13}{@{}>{\hspre}l<{\hspost}@{}}%
\column{E}{@{}>{\hspre}l<{\hspost}@{}}%
\>[B]{}\mathbf{instance}\;\Conid{Safe}\;(\Conid{IntT}\;(\Conid{Pos}\;\Conid{Z}),\Conid{IntT}\;(\Conid{Pos}\;\Conid{Z}))\;\Varid{b}{}\<[E]%
\\[\blanklineskip]%
\>[B]{}\mathbf{instance}\;({}\<[13]%
\>[13]{}\Conid{Safe}\;(\Conid{IntT}\;(\Conid{Pred}\;\Varid{n}),\Conid{IntT}\;\Varid{n'})\;\Varid{b},\Conid{Safe}\;(\Conid{IntT}\;\Varid{n},\Conid{IntT}\;(\Conid{Pred}\;\Varid{n'}))\;\Varid{b},{}\<[E]%
\\
\>[13]{}\Conid{InBoundary}\;(\Conid{IntT}\;\Varid{n},\Conid{IntT}\;\Varid{n'})\;\Varid{b})\Rightarrow \Conid{Safe}\;(\Conid{IntT}\;\Varid{n},\Conid{IntT}\;\Varid{n'})\;\Varid{b}{}\<[E]%
\ColumnHook
\end{hscode}\resethooks
Again, the first instance states that a zero-relative index is 
safe. The second instance states that a relative index is safe if it
is a member of the boundary regions for a grid and if its predecessors
in both dimensions are safe. Appendix \ref{proof} provides a
proof-sketch of the soundness of enforcing safe indexing via \ensuremath{\Conid{Safe}}.

The Laplace operator of the introductory example is desugared by the
\ttt{fun} macro into the code shown in \myfigref{laplace_desugar_code} with
the type shown in \myfigref{laplace_desugar_type}.
The \ensuremath{\Conid{Safe}} type constraints of \ensuremath{\Varid{laplace}} thus encode its relative access
pattern and constrain the grid's boundary regions such that the grid
\emph{at least} satisfies the relative indices of the grid pattern.
\begin{figure}[ht]
\vspace{-1.5em}
\hspace{-2.6em}
\subfigure[Code]{
\begin{minipage}{0.5\linewidth}
\begin{hscode}\SaveRestoreHook
\column{B}{@{}>{\hspre}l<{\hspost}@{}}%
\column{14}{@{}>{\hspre}l<{\hspost}@{}}%
\column{20}{@{}>{\hspre}l<{\hspost}@{}}%
\column{E}{@{}>{\hspre}l<{\hspost}@{}}%
\>[B]{}\Varid{laplace}\;\Varid{g}\mathrel{=}{}\<[14]%
\>[14]{}\,\;{}\<[20]%
\>[20]{}(\Varid{index2D}\;(\Conid{Neg}\;(\Conid{S}\;\Conid{Z}),\Conid{Pos}\;\Conid{Z})\;(\mathbin{-}\mathrm{1},\mathrm{0})\;\Varid{g}){}\<[E]%
\\
\>[14]{}\,\mathbin{+}{}\<[20]%
\>[20]{}(\Varid{index2D}\;(\Conid{Pos}\;(\Conid{S}\;\Conid{Z}),\Conid{Pos}\;\Conid{Z})\;(\mathrm{1},\mathrm{0})\;\Varid{g}){}\<[E]%
\\
\>[14]{}\,\mathbin{+}{}\<[20]%
\>[20]{}(\Varid{index2D}\;(\Conid{Pos}\;\Conid{Z},\Conid{Pos}\;(\Conid{S}\;\Conid{Z}))\;(\mathrm{0},\mathrm{1})\;\Varid{g}){}\<[E]%
\\
\>[14]{}\,\mathbin{+}{}\<[20]%
\>[20]{}(\Varid{index2D}\;(\Conid{Pos}\;\Conid{Z},\Conid{Neg}\;(\Conid{S}\;\Conid{Z}))\;(\mathrm{0},\mathbin{-}\mathrm{1})\;\Varid{g}){}\<[E]%
\\
\>[14]{}\,\mathbin{-}{}\<[20]%
\>[20]{}\mathrm{4}\mathbin{*}(\Varid{index2D}\;(\Conid{Pos}\;\Conid{Z},\Conid{Pos}\;\Conid{Z})\;(\mathrm{0},\mathrm{0})\;\Varid{g}){}\<[E]%
\\
\>[14]{}\,{}\<[E]%
\ColumnHook
\end{hscode}\resethooks
\end{minipage}
\label{laplace_desugar_code}
}
\hspace{1.4em}
\subfigure[Type]{
\centering
\begin{minipage}{0.5\linewidth}
\begin{hscode}\SaveRestoreHook
\column{B}{@{}>{\hspre}l<{\hspost}@{}}%
\column{4}{@{}>{\hspre}l<{\hspost}@{}}%
\column{E}{@{}>{\hspre}l<{\hspost}@{}}%
\>[B]{}({}\<[4]%
\>[4]{}\Conid{Safe}\;(\Conid{IntT}\;(\Conid{Neg}\;(\Conid{S}\;\Conid{Z})),\Conid{IntT}\;(\Conid{Pos}\;\Conid{Z}))\;\Varid{b},{}\<[E]%
\\
\>[4]{}\Conid{Safe}\;(\Conid{IntT}\;(\Conid{Pos}\;(\Conid{S}\;\Conid{Z})),\Conid{IntT}\;(\Conid{Pos}\;\Conid{Z}))\;\Varid{b},{}\<[E]%
\\
\>[4]{}\Conid{Safe}\;(\Conid{IntT}\;(\Conid{Pos}\;\Conid{Z}),\Conid{IntT}\;(\Conid{Pos}\;(\Conid{S}\;\Conid{Z})))\;\Varid{b},{}\<[E]%
\\
\>[4]{}\Conid{Safe}\;(\Conid{IntT}\;(\Conid{Pos}\;\Conid{Z}),\Conid{IntT}\;(\Conid{Neg}\;(\Conid{S}\;\Conid{Z})))\;\Varid{b},{}\<[E]%
\\
\>[4]{}\Conid{Safe}\;(\Conid{IntT}\;(\Conid{Pos}\;\Conid{Z}),\Conid{IntT}\;(\Conid{Pos}\;\Conid{Z}))\;\Varid{b},\Conid{Num}\;\Varid{a}){}\<[E]%
\\
\>[B]{}\Rightarrow \Conid{Grid}\;(\Conid{Dim}\;\Varid{d}\,{:}{^{*}}\,\Conid{Dim}\;\Varid{d'})\;\Varid{b}\;\Varid{dyn}\;\Varid{a}\to \Varid{a}{}\<[E]%
\ColumnHook
\end{hscode}\resethooks
\end{minipage}
\label{laplace_desugar_type}
}
\caption{Desugared Laplace example}
\vspace{-0.5em}
\end{figure}

\subsection{Applying Stencil Functions to Grids}
\label{stencil}

In the Laplace example, the \ensuremath{\Varid{laplace}} stencil function is
applied to a grid using \ensuremath{\Varid{runA}}. The \ensuremath{\Varid{runA}} function is
overloaded on the type parameter of a grid which encodes
whether the grid's boundary is static or dynamic, and is defined by the type class:
\begin{hscode}\SaveRestoreHook
\column{B}{@{}>{\hspre}l<{\hspost}@{}}%
\column{6}{@{}>{\hspre}l<{\hspost}@{}}%
\column{E}{@{}>{\hspre}l<{\hspost}@{}}%
\>[B]{}\mathbf{class}\;\Conid{RunGridA}\;\Varid{dyn}\;\mathbf{where}{}\<[E]%
\\
\>[B]{}\hsindent{6}{}\<[6]%
\>[6]{}\Varid{runA}\mathbin{::}(\Conid{Grid}\;\Varid{d}\;\Varid{b}\;\Varid{dyn}\;\Varid{a}\to \Varid{a})\to \Conid{Grid}\;\Varid{d}\;\Varid{b}\;\Varid{dyn}\;\Varid{a}\to \Conid{Grid}\;\Varid{d}\;\Varid{b}\;\Varid{dyn}\;\Varid{a}{}\<[E]%
\ColumnHook
\end{hscode}\resethooks
The instance for \ensuremath{\Conid{Dynamic}} applies the stencil function to a grid and
then, from the \ensuremath{\Conid{BoundaryList}} of the parameter grid, recomputes the
boundary elements. The instance for \ensuremath{\Conid{Static}} applies
the stencil function but preserves the boundary elements as
they do not require recomputation from the updated grid values.

A non-overloaded operation, called \ensuremath{\Varid{run}}, is also provided which
applies stencil functions which change the element type of a
grid, with type:
\begin{hscode}\SaveRestoreHook
\column{B}{@{}>{\hspre}l<{\hspost}@{}}%
\column{E}{@{}>{\hspre}l<{\hspost}@{}}%
\>[B]{}\Varid{run}\mathbin{::}(\Conid{IArray}\;\Conid{UArray}\;\Varid{y})\Rightarrow (\Conid{Grid}\;\Varid{d}\;\Varid{b}\;\Varid{dyn}\;\Varid{x}\to \Varid{y})\to \Conid{Grid}\;\Varid{d}\;\Varid{b}\;\Varid{dyn}\;\Varid{x}\to \Conid{Grid}\;\Varid{d}\;\Conid{Nil}\;\Conid{Static}\;\Varid{y}{}\<[E]%
\ColumnHook
\end{hscode}\resethooks
The element type of the parameter grid differs from the
element type of the return grid therefore the boundary information of the
parameter grid must be discarded as the boundary elements are of type
\ensuremath{\Varid{x}} but the grid elements are now of type \ensuremath{\Varid{y}}. The
boundary information on the return grid is thus empty i.e. $Nil$.

As a theoretical aside: Ypnos grid's are \emph{comonadic}
structures, where \ensuremath{\Varid{run}} and \ensuremath{\Varid{runA}} provide the \emph{extension}
operation. The structuring of Ypnos by \emph{comonads}
was discussed in our earlier paper \cite{ref-ypnos}.

\section{Results \& Analysis}
\label{results}

We provide a quantitative comparison of Ypnos versus Haskell with
two benchmark programs of the two-dimensional Laplace operator and
a Laplacian of Gaussian operator, which has a
larger access pattern.
All experiments were performed on an Intel Core 2 Duo 2Ghz, with 2GB
DDR3 memory, under Mac OS X version 10.5.8 using GHC
7.0.1\footnote{\url{http://www.haskell.org/ghc/download_ghc_7_0_1}}.
All code was compiled using the \ttt{-O2} flag for the
highest level of ``non-dangerous'' (i.e. no-worse performance)
optimisations. The Ypnos library is imported into a
program via an \ttt{\#include} statement, allowing whole-program
compilation for better optimisation.

All experiments are performed on a grid of size $512 \times
512$. The execution time of the whole program is measured for one
iteration of the stencil computation and for 101 iterations of the
stencil computation. The mean time per iteration is computed by the
difference of these two results divided by a $100$. The mean of ten
runs is taken with all results given to four significant figures.
\paragraph{Laplace operator} This benchmark uses the Laplace example
of Section \ref{ypnos}.


\begin{center}
\begin{tabular}{l|llll}
& Haskell & Ypnos \\ \hline
1 iteration (s) & 3.957 & 5.179 \\
101 iterations (s) & 6.297 & 7.640 \\
Mean time per iteration (s) & 0.0234 & 0.0246 \\
\end{tabular}
\end{center}
\begin{equation*}
\text{Ratio of mean time per iteration for Ypnos over Haskell = }
\frac{(7.640-5.179)/100}{(6.297-3.957)/100} = \frac{0.0246}{0.0234} \cong 1.051
\end{equation*}
i.e. the Ypnos implementation is approximately $5\%$ slower per
iteration than the Haskell implementation.
%

\paragraph{Laplacian of Gaussian}

The Laplacian of Gaussian operation combines Laplace and Gaussian
convolutions. We use here a 5$\times$5 Laplacian of Gaussian convolution operator with
coefficients \cite{machine-vision}:
{\footnotesize{
\[
\left[\begin{array}{ccccc}
0 & 0 & -1 & 0 & 0 \\
0 & -1 & -2 & -1 & 0 \\
-1 & -2 & 16 & -2 & -1 \\
0 & -1 & -2 & -1 & 0 \\
0 & 0 & -1 & 0 & 0
\end{array}
\right]
\]
}}
\begin{center}
\begin{tabular}{l|lll}
& Haskell & Ypnos \\ \hline
1 iteration (s) & 3.962 & 5.195 \\
101 iterations (s) & 7.521 & 8.662 \\
Mean time per iteration (s) & 0.0356 & 0.0347
\end{tabular}
\end{center}

\begin{equation*}
\text{Ratio of mean time per iteration for Ypnos over Haskell = }
\frac{0.0347}{0.0356} \cong 0.975
\end{equation*}
i.e. the Ypnos implementation is roughly $3\%$ faster per iteration
than the Haskell implementation.

\paragraph{Discussion}

In terms of whole program execution time, Ypnos performance is worse
than Haskell, mostly due to the overhead of the general grid and
boundary types and the cost of constructing boundary elements from
boundary definitions. However, per iteration we see that Ypnos
performance is comparable to that of Haskell. In the Laplace example,
performance is slightly worse, where the benefits of bounds-check
elimination do not offset any overhead incurred by the Ypnos
infrastructure. Given that safe indexing has an overhead of roughly
$20-25\%$ over unsafe indexing, for the two-dimensional Laplace
operator on a $512 \times 512$ grid (see Appendix
\ref{compare-safe-unsafe}), then the
overhead of Ypnos in the Laplace experiment is roughly $25-30\%$ per
iteration. For the Laplace of Gaussians example performance is however
slightly better (roughly $3\%$). Given intensive data access the
benefits of bounds-check elimination begin to overshadow any overheads
from Ypnos. Given a larger program with more stencil computations we
conjecture that performance of Ypnos could considerably surpass that
of Haskell's. 

Whilst Ypnos incurs some overhead it provides the additional benefits
of static correctness guarantees, the comparable ease of writing
stencil computations compared to Haskell or some other language, and
the generality of extending easily to higher-dimensional problems. We
are currently in the process of trying larger benchmarks in Ypnos, such as a
Navier-Stokes fluid simulator, and implementing parallel
implementations of the stencil-function application operations.





\section{Related Work}
\label{related-work}

The benefit to performance from eliminating or reducing bounds checking is well
known, first appearing in \cite{markstein1982optimization} where
various optimisations were used to reduce the cost of bounds checking
in loops. There has been research on analysis and transformation
techniques for eliminating bounds checks in various languages, such as
in Java \cite{bodik2000abcd, wurthinger2007array}.
The idea of using types to ensure safety of array operations thus
allowing array bounds checking to be eliminated is also not new. In
Xi's thesis, \emph{Dependent Types in Practical Programming},
dependently-typed array operations enforce safety of array
indexing to allow bounds-check elimination in a general setting 
\cite{xi1998dependent-thesis}. The approach in Ypnos is similar,
but using the more lightweight approach of GADTs.
Recent work by Swierstra and Altenkirch has modelled distributed array
access in the dependently-typed language Agda
\cite{swierstra2008dependent} with safe indexing. 
Sheard et. al provide a good discussion of the relation between GADTs
and dependent-types \cite{sheardgadts+}. 

The approach of using GADTs, type families, and classes to encode
invariants of a DSL is not novel. For example, Guillemette and Monnier
implement a compiler for System F inside Haskell using GADTs for a
typed-representation of System F terms
\cite{guillemette2008type}. Various transformations on the typed-terms
are mechanised via type families, such as substitution and CPS
conversion. Our approach is similar,
although we do not use GADTs to construct typed syntax trees, instead
our embedding is \emph{shallow}, where Ypnos terms are
either Haskell expressions or are desugared into Haskell expressions, in
the case of grid patterns and boundary definitions. Sackman and
Eisenbach use GADTs along with type classes to encode and enforce
invariants of DSLs at the type-level \cite{sackman2009safely}.

In \cite{kiselyov2004strongly}, type classes and regular ADTs are used
to encode type-level natural numbers and lists, instead of GADTs.
Ypnos could have used a similar scheme but we found GADTs to be more
concise, clearer, and easier to write functions over. The class/ADT
approach requires any function on the data types to be written via a
type class with each pattern matching case encoded as a class instance.

We know of two approaches to stencil programming in Haskell that are
closely related to Ypnos: PASTHA \cite{pastha} and Repa \cite{repa1,
  repa2}.

Repa provides a library of higher-order functions for programming
parallel regular, multi-dimensional arrays in Haskell
\cite{repa1}. Repa allows \emph{shape
  polymorphic} array operations via a type-level representation of
shape, or dimensionality, that is somewhat similar to the
dimensionality types of Ypnos, although Repa does not name its dimensions (the
reason for named dimensions in Ypnos is discussed in Section \ref{slices}).

Latest work on Repa \cite{repa2} provides a data type for describing 
stencils in terms of stencil size and a function from indices within
the stencil's range to values. The current implementation allows a
constant or wrapped boundary to be specified which permits a
bounds-check free implementation of a stencil application function. 
%
Ypnos allows more fine-grained boundary behaviour than the current
Repa implementation via its flexible boundary definitions and
type-level safety encoding. However, it seems likely Repa will be extended
with further boundary options in the future. Whilst the Repa implementation
is currently specialised to two-dimensional arrays and stencils, Ypnos
supports higher-dimensionality stencils via the grid
pattern syntax, which is also easy to read and write.
In the future, Repa may prove a useful implementation platform for
parallel Ypnos programs due to Repa's very good parallel support.

PASTHA is a similar library for parallelising stencil
computations in Haskell \cite{pastha}. Stencils in PASTHA are
described via a data structure and are restricted to relative indexing
in two spatial dimensions but can also index the elements of previous
iterations. The approach of Ypnos is much more general and provides more static
guarantees about correctness.

There are many other languages designed for fast and parallel array
programming such as Chapel \cite{chapel-stencils}, Titanium
\cite{yelick2007parallel}, and Single Assignment C (SAC) \cite{sac}.
Ypnos differs from these in that it is sufficiently
restricted to a problem domain, and sufficiently
expressive within that problem domain, that all information required
for guaranteeing safety, optimisation, and parallelisation is
decidable and available statically without the need for complex
compiler analysis and transformation. 

Titanium, a parallel dialect of Java developed for over a decade,
provides support for stencil computations via array loop constructs
(\ttt{foreach}) and operations
on indices (such as translations) \cite{yelick2007parallel}.
Bounds-checking can be eliminated in some cases via compiler analyses
and transformation. Any checks not eliminated provide runtime
out-of-bounds errors.
%
A special compilation mode allows bounds-checks to be removed entirely
\cite{yelick2007parallel}, which, whilst a useful software engineering
technique for production code, does not improve the robustness or
inherent efficiency of the language.
%


There has been much work on the optimisation of stencil computations
particularly in the context of optimising cache performance
(see for example \cite{kamil2006implicit}) and in the context of parallel
implementations \cite{datta2008stencil,
  krishnamoorthy2007effective}. Ypnos does not currently use any such
optimisation techniques but could certainly be improved by the use of
more sophisticated
implementations of the stencil application functions. Various
optimisation techniques such a loop tiling, skewing, or specialised
data layouts, could be tuned from the symbolic information about
access patterns statically provided by grid patterns.

\section{Further Work}
\label{further-work}

The first Ypnos paper \cite{ref-ypnos} discussed Ypnos's \emph{implementation
  agnosticism}, where the implementation of the language constructs is
parameterisable, allowing implementations tailored to different
architectures and execution strategies. In this paper we did not
discuss back-end parameterisability, but instead used a single
implementation for unboxed arrays. Currently we are in the
process of generalising the typed approach of this paper to 
a parameterisable back-end. We discuss briefly here other areas of further work.

\paragraph{Slices}
\label{slices}

One of the main reasons that Ypnos was designed with named dimensions,
e.g. \ensuremath{\Conid{X}}, \ensuremath{\Conid{Y}}, etc. was to ease multi-dimensional
programming. Named dimensions are akin to using records to
provide \emph{named} rather than \emph{positional} parameters to a
function or procedure in a language.

Further work is to add array \emph{slicing} operations to Ypnos,
for which the named dimension are particularly useful.
Slicing can be added to Ypnos by varying the behaviour of grid
patterns depending on the dimensionality of the grid matched upon. For
example, consider the hypothetical type of the following function
which uses a one-dimensional grid pattern on the \ensuremath{\Conid{X}} dimension:
\begin{Verbatim}[frame=none]
foo :: Grid (X :* d) b dyn a -> (Grid d b dyn a, Grid d b dyn a, Grid d b dyn a)
foo = [fun| X:| l {\at}c r | -> (l, c, r) |]
\end{Verbatim}
Applying \ttt{foo} to a grid of type \ensuremath{\Conid{Grid}\;(\Conid{Dim}\;\Conid{X}\,{:}{^{*}}\,\Conid{Dim}\;\Conid{Y})\;\Varid{b}\;\Varid{dyn}\;\Varid{a}} would yield a triple of successive grid \emph{slices} in the \ensuremath{\Conid{Y}}
dimension; applying \ttt{foo} to a one-dimensional grid of type \ensuremath{\Conid{Grid}\;\Conid{X}\;\Varid{b}\;\Varid{dyn}\;\Varid{a}} would yield a triple of single \ensuremath{\Varid{a}} values, requiring a unit for
the dimension tensor e.g. \ensuremath{\Conid{NoDim}} such that:
\[
\ensuremath{\Conid{NoDim}} \; \ensuremath{\,{:}{^{*}}\,} \, \ensuremath{\Conid{Dim}\;\Conid{X}} \; = \; \ensuremath{\Conid{Dim}\;\Conid{X}} \; = \; \ensuremath{\Conid{Dim}\;\Conid{X}} \; \ensuremath{\,{:}{^{*}}\,} \,
\ensuremath{\Conid{NoDim}}
\]
Additionally, \ensuremath{\Conid{Grid}\;\Conid{NoDim}\;\Varid{b}\;\Varid{dyn}\;\Varid{a}\mathrel{=}\Varid{a}}. Named dimensions allow \ttt{foo} to be
applied to a grid of the type \ensuremath{\Conid{Grid}\;(\Conid{Dim}\;\Conid{Y}\,{:}{^{*}}\,\Conid{Dim}\;\Conid{X})\;\Varid{b}\;\Varid{dyn}\;\Varid{a}} yielding the
same slicing on the \ensuremath{\Conid{X}} dimension.


Grid slicing eases higher-dimensional programming as grid
patterns need not match all dimensions at once, but could
slice a grid into lower-dimensional subgrids to be operated on by
further stencil functions.



\vspace{-1em}

\paragraph{Inductive Tuples}
\label{inductive-tuples}

Ypnos is not fully general in the dimensionality of its grids due to
the non-inductive nature of Haskell tuple types. The implementation
specifies operations usually up to four dimensions, however it is
conceivable that some applications might require
higher-dimensionality.

Inductively defined tuples, essentially lists, could be used as
indices to provide a fully general implementation, however such a
scheme is inefficient. In preliminary experiments we found that
indexing with even just a two-element list took roughly four times as
long per iteration than indexing with a tuple of two elements for a
two-dimensional Laplace operation on a $512\times512$ grid (see Appendix
\ref{tuples} for details).

An ideal system would use an inductively defined implementation with a
compiler transformation mapping inductive lists to fixed-size tuple
types in the compiled Haskell code.

\vspace{-1em}

\paragraph{Errors}
\label{errors}

A well-known problem with building embedded DSLs in statically-typed
languages is the generation of type errors which are lengthy and
confusing for the end-user. In Ypnos, if a stencil function is applied
to a grid which does not have sufficient boundary regions for safe
indexing then a type error is produced. Unfortunately
this type error is long, confusing, and intimidating.

A potential remedy for such problems could be to allow error messages
from a compiler to be passed to a pre-processor function before they
are shown to a user. For example, GHC could be extended to allow a
function to be specified which, in the event of a type-error, is passed
an AST of the error message, returning a more helpful message to the
end-user.


\section{Conclusion and Reflection}
\label{conclusion}


In our work on Ypnos our slogan has been: to develop languages to improve \emph{the four
``Rs''} of programs: \emph{reading}, \emph{(w)riting}, \emph{reasoning}, and
\emph{running}. That is, to improve the readability of programs in the
language, to improve the ease with which a program can be expressed in
the language, to increase the ability to reason about programs,
and to provide an efficient implementation. In a general purpose
language it is harder to excel in all four areas at once.
Well-designed DSLs however can simultaneously provide problem specific
expressivity, improve programmer productivity, provide stronger
reasoning, and provide more appropriate optimisation than
general-purpose languages \cite{edsl-position}. 


In Ypnos, problem specific expressivity is provided by grid
patterns and boundary definitions. General array indexing
constructions, as found in most
languages, have low information-content with respect
to program properties. Program properties relating to indexing must
often be inferred via analysis, which may be undecidable in general.
Grid patterns and boundary definitions provide a
restricted but information-rich method of indexing. We believe that
Ypnos's syntax eases \emph{reading} and \emph{writing} of stencil computations,
although this is difficult to measure objectively. It is however easier to
quantify the effect on \emph{reasoning} and \emph{running}. This paper has shown
that grid patterns and boundary regions provide enough information to
statically prove safety of Ypnos programs, which also permits
optimisation via bounds-check elimination. More complex
optimisations, such as tiling, skewing,
and cache sensitive data-layout, could be performed since grid patterns provide
decidable static information about array access.

Other embedded DSLs have used the type system of the host language to
enforce invariants of the DSL (of which a few were mentioned in
Section \ref{related-work}). 
In our approach GADTs were used to lift data-level information to the
type-level; type families were used 
to manipulate and compose such information; and type classes were used to
enforce language properties/invariants based on this information.

One lesson learnt was in leveraging the semantics of type classes.
A first attempt at encoding safety for Ypnos in Haskell used
a special stencil data-type for grid patterns that included all relative
index information in its type. Safety was then enforced by unification
between this stencil type and a boundary type. However, since there
is no particular order in which boundary definitions should be
defined, a type-level normalisation algorithm, essentially a kind of
sorting, was required for unification of boundary and stencil
types. This approach was extremely complicated and error-prone. Type
class constraints proved to be much more useful in the end, as they comprise a
conjunction of constraints that is implicitly associative and
commutative, thus ordering is not relevant. Safety in Ypnos is
fortunately easily expressed via a conjunction of individual safety
constraints (as shown in Appendix \ref{proof}). Encoding disjunctions
of invariants would be more difficult and is not something we have
experimented with or required yet.
%

Another lesson learnt is that a working type-level encoding of a
language invariant does not imply soundness of the encoding.
For a time we had a type-level encoding of safety which appeared correct;
it was simple and elegant, rejected the programs we
thought it should reject, and accepted the programs we thought it
should accept. Unfortunately it was unsound, accepting
some programs that should have been rejected. Overconfidence in
the infrastructure meant it was a while before this unsoundness was
discovered. Writing a proof of soundness beforehand is preferable,
and may even give an indication of how to implement a type-level
encoding, as does the soundness proof-sketch in Appendix \ref{proof}.



There are certainly many more areas of programming that could benefit
from DSLs. \emph{Structured grids}, or \emph{stencil computations}, as
discussed in this paper, are amongst a number of programming
patterns that have been identified as key motifs in parallel
programming \cite{berkeley06}. We have already begun considering
extensions to Ypnos to support \emph{unstructured grids}, also called
\emph{meshes}. We would be interested to see if other restricted
syntactic forms such as grid patterns could be invented for other
programming patterns.






\section*{Acknowledgments}

Ypnos began as joint work with Max Bolingbroke \cite{ref-ypnos}.
His early ideas and contributions still pervade the current
instantiation of Ypnos. Thanks are due to the anonymous reviewers for
their useful feedback, to Ian McDonnell for his comments and
help on the image processing related aspects of this work, and to Ben
Lippmeier, Robin Message, and Tomas Petricek for insightful comments and feedback.

\vspace{-1em}

\bibliographystyle{eptcs}
\bibliography{references}

\appendix

\vspace{-1.8em}

\section{Supporting Experiments}
\label{experiments}

Each experiment is set up and measured as the experiments in Section \ref{results}.

\vspace{-0.8em}

\paragraph{Comparing Safe \& Unsafe Indexing}
\label{compare-safe-unsafe}

A 2D-Laplace operation was applied to a
$512\times{512}$ image using both safe and unsafe memory access to test the
effect on execution performance. The results were:

\begin{tabular}{l|ll}
& Safe indexing & Unsafe indexing \\ \hline
1 iteration (s) & 3.982 & 3.952 \\
101 iterations (s) & 6.328 & 5.865 \\
Mean time per iteration (s) & 0.0234 & 0.0191
\end{tabular} \\

\noindent
The overhead of safe indexing is approximately
$\frac{0.0234}{0.0191} \cong 1.23$ i.e. approximately $20-25\%$.

\vspace{-0.2em}

\paragraph{Comparing Tuple Indices \& Inductively-defined
  Indices}
\label{tuples}

A 2D-Laplace operation was applied to a $512\times{512}$ image using
an array indexed by a pair of \ensuremath{\Conid{Int}}s and an array indexed by an
inductively defined \ensuremath{\Conid{Int}} list structure of length two.

\begin{tabular}{l|ll}
& Tuples & Lists \\ \hline
1 iteration (s) & 3.982 & 4.133 \\
101 iterations (s) & 6.328 & 13.941 \\
Mean time per iteration (s) & 0.0234 & 0.0981
\end{tabular} \\

\noindent
$\frac{0.0981}{0.0234} \cong 4.19$, therefore the overhead of the list
index is considerable at approximately $300\%$.

\section{Haskell and GHC Type-System Features}
\label{haskell-types}


\paragraph{Type Classes}

\emph{Type classes} in Haskell provide a mechanism for overloading, also
known as \emph{ad-hoc polymorphism} or \emph{type-indexed
functions} \cite{hall1996type}. Consider the equality operation (\ensuremath{{=}{=}}). Many types have a
well-defined notion of equality, thus an overloaded \ensuremath{({=}{=})} operator is
useful, where the argument type determines the actual equality
operation used. The equality type class is defined:
\begin{hscode}\SaveRestoreHook
\column{B}{@{}>{\hspre}l<{\hspost}@{}}%
\column{4}{@{}>{\hspre}l<{\hspost}@{}}%
\column{E}{@{}>{\hspre}l<{\hspost}@{}}%
\>[B]{}\mathbf{class}\;\Conid{Eq}\;\Varid{a}\;\mathbf{where}{}\<[E]%
\\
\>[B]{}\hsindent{4}{}\<[4]%
\>[4]{}({=}{=})\mathbin{::}\Varid{a}\to \Varid{a}\to \Conid{Bool}{}\<[E]%
\ColumnHook
\end{hscode}\resethooks
A type is declared a member of a class by an \emph{instance} definition,
which provides definitions of the class functions for a particular type. For example:
\begin{hscode}\SaveRestoreHook
\column{B}{@{}>{\hspre}l<{\hspost}@{}}%
\column{4}{@{}>{\hspre}l<{\hspost}@{}}%
\column{E}{@{}>{\hspre}l<{\hspost}@{}}%
\>[B]{}\mathbf{instance}\;\Conid{Eq}\;\Conid{Int}\;\mathbf{where}{}\<[E]%
\\
\>[B]{}\hsindent{4}{}\<[4]%
\>[4]{}\Varid{x}{=}{=}\Varid{y}\mathrel{=}\Varid{eqInt}\;\Varid{x}\;\Varid{y}{}\<[E]%
\ColumnHook
\end{hscode}\resethooks
where \ensuremath{\Varid{eqInt}} is the built-in equality operation for \ensuremath{\Conid{Int}}. By
this instance definition, \ensuremath{\Conid{Int}} is a member of \ensuremath{\Conid{Eq}}.

Usage of \ensuremath{({=}{=})} on polymorphically-typed arguments imposes a \emph{type-class
  constraint} on the polymorphic type of the arguments.
Type-class constraints enforce that a type is an instance of a
class and are written on the left-hand side of a type, preceding an
arrow $\Rightarrow$. For example, the following function:
\begin{hscode}\SaveRestoreHook
\column{B}{@{}>{\hspre}l<{\hspost}@{}}%
\column{E}{@{}>{\hspre}l<{\hspost}@{}}%
\>[B]{}\Varid{f}\;\Varid{x}\;\Varid{y}\mathrel{=}\Varid{x}{=}{=}\Varid{y}{}\<[E]%
\ColumnHook
\end{hscode}\resethooks
has type \ensuremath{\Varid{f}\mathbin{::}\Conid{Eq}\;\Varid{a}\Rightarrow \Varid{a}\to \Varid{a}\to \Conid{Bool}}, where \ensuremath{\Conid{Eq}\;\Varid{a}} is the type-class
constraint that \ensuremath{\Varid{a}} be a member of \ensuremath{\Conid{Eq}}.


\paragraph{GADTs}

\emph{Generalised algebraic data types}, or GADTs,
\cite{jones2004wobbly, peyton2006simple} have become a powerful
technique for dependent-type like programming in Haskell.
Consider the following algebraic data type in Haskell which encodes a
simple term calculus of products and projections over some type:
\begin{hscode}\SaveRestoreHook
\column{B}{@{}>{\hspre}l<{\hspost}@{}}%
\column{E}{@{}>{\hspre}l<{\hspost}@{}}%
\>[B]{}\mathbf{data}\;\Conid{Term}\;\Varid{a}\mathrel{=}\Conid{Pair}\;(\Conid{Term}\;\Varid{a})\;(\Conid{Term}\;\Varid{a})\mid \Conid{Fst}\;(\Conid{Term}\;\Varid{a})\mid \Conid{Snd}\;(\Conid{Term}\;\Varid{a})\mid \Conid{Val}\;\Varid{a}{}\<[E]%
\ColumnHook
\end{hscode}\resethooks
The ADT is polymorphic in the type $a$ and is recursive. For each
\emph{tag} in a the definition of an ADT (e.g. \ensuremath{\Conid{Pair}}, \ensuremath{\Conid{Fst}}, etc.) a
constructor is generated; e.g. for the first two data constructors of \ensuremath{\Conid{Term}}:
\begin{hscode}\SaveRestoreHook
\column{B}{@{}>{\hspre}l<{\hspost}@{}}%
\column{E}{@{}>{\hspre}l<{\hspost}@{}}%
\>[B]{}\Conid{Pair}\mathbin{::}\Conid{Term}\;\Varid{a}\to \Conid{Term}\;\Varid{a}\to \Conid{Term}\;\Varid{a}{}\<[E]%
\\
\>[B]{}\Conid{Fst}\mathbin{::}\Conid{Term}\;\Varid{a}\to \Conid{Term}\;\Varid{a}{}\<[E]%
\ColumnHook
\end{hscode}\resethooks
Thus, every constructor returns a value of type \ensuremath{\Conid{Term}\;\Varid{a}}. 
We can define
another ADT of values \ensuremath{\Conid{Val}} and write an evaluator from \ensuremath{\Conid{Term}} to \ensuremath{\Conid{Val}}:
\begin{hscode}\SaveRestoreHook
\column{B}{@{}>{\hspre}l<{\hspost}@{}}%
\column{18}{@{}>{\hspre}l<{\hspost}@{}}%
\column{38}{@{}>{\hspre}l<{\hspost}@{}}%
\column{51}{@{}>{\hspre}c<{\hspost}@{}}%
\column{51E}{@{}l@{}}%
\column{55}{@{}>{\hspre}l<{\hspost}@{}}%
\column{64}{@{}>{\hspre}c<{\hspost}@{}}%
\column{64E}{@{}l@{}}%
\column{68}{@{}>{\hspre}l<{\hspost}@{}}%
\column{E}{@{}>{\hspre}l<{\hspost}@{}}%
\>[B]{}\mathbf{data}\;\Conid{Val}\;\Varid{a}\mathrel{=}\Conid{ValPair}\;(\Conid{Val}\;\Varid{a})\;(\Conid{Val}\;\Varid{a})\mid \Conid{ValConst}\;\Varid{a}{}\<[E]%
\\[\blanklineskip]%
\>[B]{}\Varid{eval}\mathbin{::}\Conid{Term}\;\Varid{a}\to \Conid{Val}\;\Varid{a}{}\<[E]%
\\
\>[B]{}\Varid{eval}\;(\Conid{Pair}\;\Varid{x}\;\Varid{y}){}\<[18]%
\>[18]{}\mathrel{=}\Conid{ValPair}\;(\Varid{eval}\;\Varid{x})\;(\Varid{eval}\;\Varid{y}){}\<[E]%
\\
\>[B]{}\Varid{eval}\;(\Conid{Fst}\;\Varid{x}){}\<[18]%
\>[18]{}\mathrel{=}\mathbf{case}\;(\Varid{eval}\;\Varid{x})\;\mathbf{of}\;{}\<[38]%
\>[38]{}\Conid{ValPair}\;\Varid{a}\;\Varid{b}{}\<[51]%
\>[51]{}\to {}\<[51E]%
\>[55]{}\Varid{a};\,\;\anonymous {}\<[64]%
\>[64]{}\to {}\<[64E]%
\>[68]{}\Varid{error}\;\text{\tt \char34 whoops\char34}{}\<[E]%
\\
\>[B]{}\Varid{eval}\;(\Conid{Snd}\;\Varid{x}){}\<[18]%
\>[18]{}\mathrel{=}\mathbf{case}\;(\Varid{eval}\;\Varid{x})\;\mathbf{of}\;{}\<[38]%
\>[38]{}\Conid{ValPair}\;\Varid{a}\;\Varid{b}{}\<[51]%
\>[51]{}\to {}\<[51E]%
\>[55]{}\Varid{b};\,\;\anonymous {}\<[64]%
\>[64]{}\to {}\<[64E]%
\>[68]{}\Varid{error}\;\text{\tt \char34 whoops\char34}{}\<[E]%
\\
\>[B]{}\Varid{eval}\;(\Conid{Val}\;\Varid{x}){}\<[18]%
\>[18]{}\mathrel{=}\Conid{ValConst}\;\Varid{x}{}\<[E]%
\ColumnHook
\end{hscode}\resethooks
Unfortunately, the \ensuremath{\Conid{Term}} ADT allows non-sensical terms
in the \ensuremath{\Conid{Term}} calculus to be constructed, such as \ensuremath{\Conid{Fst}\;(\Conid{Val}\;\mathrm{1})}, therefore
error handling cases must be included for \ensuremath{\Varid{eval}} on \ensuremath{\Conid{Fst}} and \ensuremath{\Conid{Snd}}.

The crucial difference between ADTs and GADTs is that GADTs specify
the full type of a data constructor, allowing the result type of the
constructor to have arbitrary type-parameters for the GADT's type
constructor. For example, \ensuremath{\Conid{Term}} can be rewritten as the following on
the left-hand side:

\begin{minipage}{0.5\linewidth}
\begin{hscode}\SaveRestoreHook
\column{B}{@{}>{\hspre}l<{\hspost}@{}}%
\column{5}{@{}>{\hspre}l<{\hspost}@{}}%
\column{11}{@{}>{\hspre}l<{\hspost}@{}}%
\column{E}{@{}>{\hspre}l<{\hspost}@{}}%
\>[B]{}\mathbf{data}\;\Conid{Term}\;\Varid{t}\;\mathbf{where}{}\<[E]%
\\
\>[B]{}\hsindent{5}{}\<[5]%
\>[5]{}\Conid{Pair}{}\<[11]%
\>[11]{}\mathbin{::}\Conid{Term}\;\Varid{a}\to \Conid{Term}\;\Varid{b}\to \Conid{Term}\;(\Varid{a},\Varid{b}){}\<[E]%
\\
\>[B]{}\hsindent{5}{}\<[5]%
\>[5]{}\Conid{Fst}{}\<[11]%
\>[11]{}\mathbin{::}\Conid{Term}\;(\Varid{a},\Varid{b})\to \Conid{Term}\;\Varid{a}{}\<[E]%
\\
\>[B]{}\hsindent{5}{}\<[5]%
\>[5]{}\Conid{Snd}{}\<[11]%
\>[11]{}\mathbin{::}\Conid{Term}\;(\Varid{a},\Varid{b})\to \Conid{Term}\;\Varid{b}{}\<[E]%
\\
\>[B]{}\hsindent{5}{}\<[5]%
\>[5]{}\Conid{Val}{}\<[11]%
\>[11]{}\mathbin{::}\Varid{a}\to \Conid{Term}\;\Varid{a}{}\<[E]%
\ColumnHook
\end{hscode}\resethooks
\end{minipage}
\begin{minipage}{0.5\linewidth}
\begin{hscode}\SaveRestoreHook
\column{B}{@{}>{\hspre}l<{\hspost}@{}}%
\column{18}{@{}>{\hspre}l<{\hspost}@{}}%
\column{E}{@{}>{\hspre}l<{\hspost}@{}}%
\>[B]{}\Varid{eval}\mathbin{::}\Conid{Term}\;\Varid{a}\to \Varid{a}{}\<[E]%
\\
\>[B]{}\Varid{eval}\;(\Conid{Pair}\;\Varid{x}\;\Varid{y}){}\<[18]%
\>[18]{}\mathrel{=}(\Varid{eval}\;\Varid{x},\Varid{eval}\;\Varid{y}){}\<[E]%
\\
\>[B]{}\Varid{eval}\;(\Conid{Fst}\;\Varid{x}){}\<[18]%
\>[18]{}\mathrel{=}\Varid{fst}\;(\Varid{eval}\;\Varid{x}){}\<[E]%
\\
\>[B]{}\Varid{eval}\;(\Conid{Snd}\;\Varid{x}){}\<[18]%
\>[18]{}\mathrel{=}\Varid{snd}\;(\Varid{eval}\;\Varid{x}){}\<[E]%
\\
\>[B]{}\Varid{eval}\;(\Conid{Val}\;\Varid{x}){}\<[18]%
\>[18]{}\mathrel{=}\Varid{x}{}\<[E]%
\ColumnHook
\end{hscode}\resethooks
\end{minipage}

The type parameter of \ensuremath{\Conid{Term}} encodes the type of a term as a 
Haskell type. Because a term type is known, \ensuremath{\Varid{eval}} can be rewritten
as on the right-hand side, where a well-typed Haskell value
is returned as opposed to the \ensuremath{\Conid{Val}} datatype encoding \ensuremath{\Conid{Term}} values used
previously. Furthermore, \ensuremath{\Varid{eval}} cannot be applied to
malformed \ensuremath{\Conid{Term}} terms, thus error handling cases are unnecessary. 
GADTs thus allow a form of lightweight dependent-typing by permitting
types to depend on data constructors.

\paragraph{Type Families}


\emph{Type families} in Haskell, also called \emph{type-indexed families of
types}, are simple type-level functions providing a limited form of computation at
the type-level, evaluated during type checking
\cite{associated-type-synonyms}. 
Type families describe a number of rewrite rules from types to types,
consisting of a \emph{head declaration} specifying the name and arity
of the type family and a number of \emph{instance declarations}
defining the rewrite rules. For example, the following type family
provides a projection function on pair types:
\begin{hscode}\SaveRestoreHook
\column{B}{@{}>{\hspre}l<{\hspost}@{}}%
\column{E}{@{}>{\hspre}l<{\hspost}@{}}%
\>[B]{}\mathbf{type}\;\textbf{family}\;\Conid{Fst}\;\Varid{t}{}\<[E]%
\\
\>[B]{}\mathbf{type}\;\mathbf{instance}\;\Conid{Fst}\;(\Varid{a},\Varid{b})\mathrel{=}\Varid{a}{}\<[E]%
\ColumnHook
\end{hscode}\resethooks
Type families are \emph{open} allowing further instances to be defined 
throughout a program or in another module. Instances of a
family are therefore \emph{unordered} thus a application of a family
to an argument does not result in ordered pattern
matching of the argument against the family's instances.
Consequently, to preserve uniqueness of typing,
type family instances must not
\emph{overlap} or at least must be \emph{confluent} i.e. if
there are two possible rewrites for a type family then the rewrites
are equivalent.
\noindent

Type families may be recursive, as long as the size of the type
parameters of the recursive call are less than the size of the type
parameters in the instance making the recursive call.
For example, the following defines an append
operation on the type-level list representation shown in Section \ref{indices}:
\begin{hscode}\SaveRestoreHook
\column{B}{@{}>{\hspre}l<{\hspost}@{}}%
\column{E}{@{}>{\hspre}l<{\hspost}@{}}%
\>[B]{}\mathbf{type}\;\textbf{family}\;\Conid{Append}\;\Varid{x}\;\Varid{y}{}\<[E]%
\\
\>[B]{}\mathbf{type}\;\mathbf{instance}\;\Conid{Append}\;\Conid{Nil}\;\Varid{z}\mathrel{=}\Varid{z}{}\<[E]%
\\
\>[B]{}\mathbf{type}\;\mathbf{instance}\;\Conid{Append}\;(\Conid{Cons}\;\Varid{x}\;\Varid{y})\;\Varid{z}\mathrel{=}\Conid{Cons}\;\Varid{x}\;(\Conid{Append}\;\Varid{y}\;\Varid{z}){}\<[E]%
\ColumnHook
\end{hscode}\resethooks

\section{Soundness Proof}
\label{proof}

The safe-indexing invariant of Ypnos is \emph{sound} if
well-typed programs cannot index undefined elements, that is,
any out-of-bounds access always has a defined value.
We provide a (semi-formal) proof of soundness here for two-dimensional grids, although
the proof generalises easily to arbitrary dimensions. 

Consider a grid of two-dimensions with finite extent $(0, 0)$ to
$(N, M)$ (exclusive) and a relative index $(i, j)$ accessed by some stencil
function. We assume for this proof that $i$ and $j$ are both positive relative
indices without loss of generality as all definitions are symmetric in
the sign of relative indices.

Let $V$ be a predicate on index ranges which denotes indices which
have a defined value. As an axiom, all indices inside the grid's
extent have a defined value, thus:
\begin{align}
\text{(axiom)} \quad & \quad V[0,N][0,M]
\label{base}
\end{align}
Let $S$ be a predicate of the safe relative indices for a
grid where $S (x, y)$ means $(x, y)$ is safe. For the positive
relative index $(i, j)$, $S$ and $V$ are related thus:
\begin{equation}
V[0, N+i][0, M+j] \Leftrightarrow S (i, j)
\label{S-V-rel}
\end{equation}
Axiom \eqref{base} is therefore equivalent to $S (0, 0)$ i.e. zero-relative
indices are always safe.

The range of $V$ can be separated into a conjunction of subranges,
thus we can express \eqref{base} as:
\begin{align}
\begin{array}{ll}
& V[0, N][0, M] \\
\wedge & V[0, N][M, M+j-1]  \\
\wedge & V[N, N+i-1][0, M] \\
\wedge & V[N, N+i][M, M+j]
\end{array} \quad \Rightarrow \quad S (i, j)
\label{split}
\end{align}
(Note: by \eqref{base} we could eliminate the first conjunct)
Since logical conjunction is \emph{involutive} (i.e. $A \wedge A = A$)
we can overlap the regions of $V$, thus we could rewrite \eqref{split} as:
\begin{align}
\begin{array}{ll}
& V[0, N+i-1][0, M+j] \\
\wedge & V[0, N+i][0, M+j-1] \\
\wedge & V[N+i, N+i][M+j, M+j]
\end{array} \quad \Rightarrow \quad S (i, j)
\label{split2}
\end{align}

Now consider another predicate of defined boundary regions where
$B(x, y)$ means that the boundary region $(x, y)$ is defined. The
boundary region predicate $B$ is related to $V$ in the following way:
\begin{align}
\begin{array}{ll}
& B (\ttt{+}x, \ttt{+}y) \Rightarrow V[N+x, N+x][M+y, M+y] \\
& B (\ttt{+}x, \ttt{-}y) \Rightarrow V[N+x, N+x][-y, -y] \\
& B (\ttt{-}x, \ttt{+}y) \Rightarrow V[-x, -x][M+y, M+y] \\
& B (\ttt{-}x, \ttt{-}y) \Rightarrow V[-x, -x][-y, -y]
\end{array}
\begin{array}{ll}
& B (\ttt{+}x, \ttt{*}) \Rightarrow V[N+x, N+x][0, M] \\
& B (\ttt{*v}, \ttt{+}y) \Rightarrow V[0, N][M+y, M+y] \\
& B (\ttt{-}x, \ttt{*v}) \Rightarrow V[-x, -x][0, M] \\
& B (\ttt{*v}, \ttt{-}y) \Rightarrow V[0, N][-y, -y]
\end{array}
\label{bound-to-v}
\end{align}

Boundary regions thus map to a single defined index e.g. $V[N+x,
N+x][N+y, N+y]$, or to a range over the whole inner extent in one 
dimension with the other dimension fixed at a single index outside
of the extent e.g. $V[N+x, N+x][0, M]$ or $V[0, N][-y, -y]$.
%

By \eqref{bound-to-v} we can replace $V[N+i, N+i][M+j,
  M+j]$ in \eqref{split2} with $B (i, j)$, thus:
\begin{align}
\begin{array}{ll}
& V[0, N+i-1][0, M+j] \\
\wedge & V[0, N+i][0, M+j-1] \\
\wedge & B(i, j)
\end{array} \quad \Rightarrow \quad S (i, j)
\label{split3}
\end{align}
We can eliminate $V$ from \eqref{split3} entirely by the
relation of $V$ to $S$ \eqref{S-V-rel}:
\begin{equation}
S (i - 1, j) \; \wedge \; S(i, j - 1) \; \wedge \; B (i, j) \; \Rightarrow \; S(i, j) 
\label{recurrence}
\end{equation}
Equation \eqref{recurrence} thus specifies a kind of recurrence
relation on \ensuremath{\Conid{S}}, which will eventually reach the axiomatic base case $S(0, 0)$.
The definition of \ensuremath{\Conid{Safe}} is exactly the predicate $S$ as defined by the
recurrence \eqref{recurrence} where \ensuremath{\Conid{InBoundary}} is the predicate $B$:
\begin{hscode}\SaveRestoreHook
\column{B}{@{}>{\hspre}l<{\hspost}@{}}%
\column{13}{@{}>{\hspre}l<{\hspost}@{}}%
\column{E}{@{}>{\hspre}l<{\hspost}@{}}%
\>[B]{}\mathbf{instance}\;({}\<[13]%
\>[13]{}\Conid{Safe}\;(\Conid{IntT}\;(\Conid{Pred}\;\Varid{n}),\Conid{IntT}\;\Varid{n'})\;\Varid{b},\Conid{Safe}\;(\Conid{IntT}\;\Varid{n},\Conid{IntT}\;(\Conid{Pred}\;\Varid{n'}))\;\Varid{b},{}\<[E]%
\\
\>[13]{}\Conid{InBoundary}\;(\Conid{IntT}\;\Varid{n},\Conid{IntT}\;\Varid{n'})\;\Varid{b})\Rightarrow \Conid{Safe}\;(\Conid{IntT}\;\Varid{n},\Conid{IntT}\;\Varid{n'})\;\Varid{b}{}\<[E]%
\ColumnHook
\end{hscode}\resethooks
The axiom \eqref{base}, $S (0, 0)$ is satisfied by the case that all zero-relative
indices are safe:
\begin{hscode}\SaveRestoreHook
\column{B}{@{}>{\hspre}l<{\hspost}@{}}%
\column{E}{@{}>{\hspre}l<{\hspost}@{}}%
\>[B]{}\mathbf{instance}\;\Conid{Safe}\;(\Conid{IntT}\;(\Conid{Pos}\;\Conid{Z}),\Conid{IntT}\;(\Conid{Pos}\;\Conid{Z}))\;\Varid{b}{}\<[E]%
\ColumnHook
\end{hscode}\resethooks
By induction \ensuremath{\Conid{Pred}} reduces relative indices towards zero, acting as
the minus operation in \eqref{recurrence}, and is symmetrical in its
treatment of negative and positive relative indices.

\ensuremath{\Conid{Safe}} thus provides a sound encoding of safe-indexing for grids. $\Box{}$



%
%
%
%
%

\end{document}